\documentclass[twoside,twocolumn,9pt]{article}
\usepackage{extsizes}
\usepackage[super,sort&compress,comma]{natbib} 
\usepackage[version=3]{mhchem}
\usepackage[left=1.5cm, right=1.5cm, top=1.785cm, bottom=2.0cm]{geometry}
\usepackage{balance}
\usepackage{mathptmx}
\usepackage{sectsty}
\usepackage{graphicx} 
\usepackage{lastpage}
\usepackage[format=plain,justification=justified,singlelinecheck=false,font={stretch=1.125,small,sf},labelfont=bf,labelsep=space]{caption}
\usepackage{float}
\usepackage{fancyhdr}
\usepackage{fnpos}
\usepackage[english]{babel}
\addto{\captionsenglish}{%
  
}
\usepackage{array}
\usepackage{droidsans}
\usepackage{charter}
\usepackage[T1]{fontenc}
\usepackage[usenames,dvipsnames]{xcolor}
\usepackage{setspace}
\usepackage[compact]{titlesec}
\usepackage{hyperref}
\usepackage{amssymb}
\usepackage{siunitx}

\usepackage{epstopdf}%This line makes .eps figures into .pdf - please comment out if not required.

\definecolor{cream}{RGB}{222,217,201}

\begin{document}

\pagestyle{fancy}
\thispagestyle{plain}
\fancypagestyle{plain}{
%%%HEADER%%%
\renewcommand{\headrulewidth}{0pt}
}
%%%END OF HEADER%%%

%%%PAGE SETUP - Please do not change any commands within this section%%%
\makeFNbottom
\makeatletter
\renewcommand\LARGE{\@setfontsize\LARGE{15pt}{17}}
\renewcommand\Large{\@setfontsize\Large{12pt}{14}}
\renewcommand\large{\@setfontsize\large{10pt}{12}}
\renewcommand\footnotesize{\@setfontsize\footnotesize{7pt}{10}}
\makeatother

\renewcommand{\thefootnote}{\fnsymbol{footnote}}
\renewcommand\footnoterule{\vspace*{1pt}% 
\color{cream}\hrule width 3.5in height 0.4pt \color{black}\vspace*{5pt}} 
\setcounter{secnumdepth}{5}

\makeatletter 
\renewcommand\@biblabel[1]{#1}            
\renewcommand\@makefntext[1]% 
{\noindent\makebox[0pt][r]{\@thefnmark\,}#1}
\makeatother 
\renewcommand{\figurename}{\small{Fig.}~}
\sectionfont{\sffamily\Large}
\subsectionfont{\normalsize}
\subsubsectionfont{\bf}
\setstretch{1.125} %In particular, please do not alter this line.
\setlength{\skip\footins}{0.8cm}
\setlength{\footnotesep}{0.25cm}
\setlength{\jot}{10pt}
\titlespacing*{\section}{0pt}{4pt}{4pt}
\titlespacing*{\subsection}{0pt}{15pt}{1pt}
%%%END OF PAGE SETUP%%%

%%%FOOTER%%%
\fancyfoot{}
\fancyfoot[LO,RE]{\vspace{-7.1pt}\includegraphics[height=9pt]{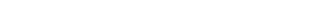}}
\fancyfoot[CO]{\vspace{-7.1pt}\hspace{13.2cm}\includegraphics{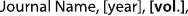}}
\fancyfoot[CE]{\vspace{-7.2pt}\hspace{-14.2cm}\includegraphics{head_foot/RF}}
\fancyfoot[RO]{\footnotesize{\sffamily{1--\pageref{LastPage} ~\textbar  \hspace{2pt}\thepage}}}
\fancyfoot[LE]{\footnotesize{\sffamily{\thepage~\textbar\hspace{3.45cm} 1--\pageref{LastPage}}}}
\fancyhead{}
\renewcommand{\headrulewidth}{0pt} 
\renewcommand{\footrulewidth}{0pt}
\setlength{\arrayrulewidth}{1pt}
\setlength{\columnsep}{6.5mm}
\setlength\bibsep{1pt}
%%%END OF FOOTER%%%

%%%FIGURE SETUP - please do not change any commands within this section%%%
\makeatletter 
\newlength{\figrulesep} 
\setlength{\figrulesep}{0.5\textfloatsep} 

\newcommand{\topfigrule}{\vspace*{-1pt}% 
\noindent{\color{cream}\rule[-\figrulesep]{\columnwidth}{1.5pt}} }

\newcommand{\botfigrule}{\vspace*{-2pt}% 
\noindent{\color{cream}\rule[\figrulesep]{\columnwidth}{1.5pt}} }

\newcommand{\dblfigrule}{\vspace*{-1pt}% 
\noindent{\color{cream}\rule[-\figrulesep]{\textwidth}{1.5pt}} }

\makeatother
%%%END OF FIGURE SETUP%%%
%%%TITLE, AUTHORS AND ABSTRACT%%%
\twocolumn[
  \begin{@twocolumnfalse}
{\includegraphics[height=30pt]{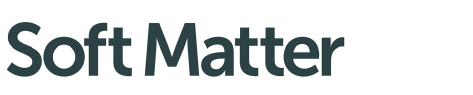}\hfill\raisebox{0pt}[0pt][0pt]{\includegraphics[height=55pt]{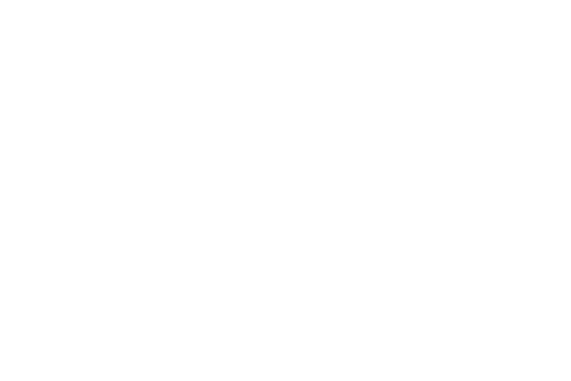}}\\[1ex]
\includegraphics[width=18.5cm]{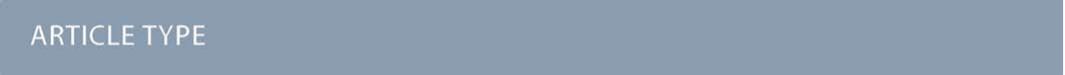}}\par
\vspace{1em}
\sffamily
\begin{tabular}{m{4.5cm} p{13.5cm} }
%%%%%%%%%%%%%%%%%%%%%%%%%%%%%%%%%%%%%%%%%%%%%%%%%%%%%%%%%%%%%%%%%%%%%%%%%%%%%%%%%%%%%%%%%%%%%%%%%%%%%%
\includegraphics{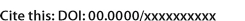} & \noindent\LARGE{\textbf{Effects of Near-Field Hydrodynamic Interactions on Bacterial Dynamics Near a Solid Surface}} \\%Article title goes here instead of the text "This is the title"
\vspace{0.3cm} & \vspace{0.3cm} \\
 & \noindent\large{Baopi Liu,$^{\ast}$\textit{$^{a,b}$} Lu Chen,\textit{$^{c}$} and Haiqin Wang~\textit{$^{d}$}} \\%Author names go here instead of "Full name", etc.
%%%%%%%%%%%%%%%%%%%%%%%%%%%%%%%%%%%%%%%%%%%%%%%%%%%%%%%%%%%%%%%%%%%%%%%%%%%%%%%%%%%%%%%%%%%%%%%%%%%%%%
\includegraphics{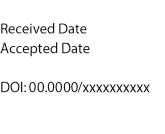} & \noindent\normalsize{Near-field hydrodynamic interactions between bacteria and no-slip solid surfaces are the main mechanism underlying surface entrapment of bacteria. In this study, we employ a chiral two-body model to simulate bacterial dynamics near the surface. The simulation results show that as bacteria approach the surface, their translational velocities and diffusion coefficients decrease. Under the combination of near-field hydrodynamic interactions and DLVO forces, bacteria reach a stable fixed point in the phase plane and follow circular trajectories at this point. In particular, bacteria with left-handed helical flagella exhibit clockwise circular motion on the surface. During this process, as the stable height increases, the near-field hydrodynamic interactions weaken. Consequently, the translational velocity of the bacteria parallel to the surface increases while the rotational velocity perpendicular to the surface decreases, collectively increasing the radius of curvature. Ultimately, our findings demonstrate that near-field hydrodynamic interactions significantly prolong the surface residence time of bacteria. Additionally, smaller stable heights further amplify this effect, resulting in longer residence times and enhanced surface entrapment.} 
%The abstrast goes here instead of the text "The abstract should be..."
%%%%%%%%%%%%%%%%%%%%%%%%%%%%%%%%%%%%%%%%%%%%%%%%%%%%%%%%%%%%%%%%%%%%%%%%%%%%%%%%%%%%%%%%%%%%%%%%%%%%%%
\end{tabular}
 \end{@twocolumnfalse} \vspace{0.6cm}
  ]
%%%END OF TITLE, AUTHORS AND ABSTRACT%%%
%%%FONT SETUP - please do not change any commands within this section
\renewcommand*\rmdefault{bch}\normalfont\upshape
\rmfamily
\section*{}
\vspace{-1cm}
%%%FOOTNOTES%%%
\footnotetext{\textit{$^{*}$~E-mail: bpliu@mail.bnu.edu.cn}}
\footnotetext{\textit{$^{a}$~School of Advanced Interdisciplinary Studies, Ningxia University, Zhongwei, 755000, China}}
\footnotetext{\textit{$^{b}$~Complex Systems Division, Beijing Computational Science Research Center, Beijing 100193, China}}
\footnotetext{\textit{$^{c}$~College of Physics, Changchun Normal University, Changchun, Jilin 130032, China}}
\footnotetext{\textit{$^{d}$~Department of Chemistry, Harvey Mudd College, Claremont, California 91711, USA}}
%%%%%%%%%%%%%%%%%%%%%%%%%%%%%%%%%%%%%%%%%%%%%%%%%%%%%%%%%%%%%%%%%%%%%%%%%%%%%%%%%%%%%%%%%%%%%%%%%%%%%%
%%%END OF FOOTNOTES%%%
%%%MAIN TEXT%%%%
%%%%%%%%%%%%%%%%%%%%%%%%%%%%%%%%%%%%%%%%%%%%%%%%%%%%%%%%%%%%%%%%%%%%%%%%%%%%%%%%%%%%%%%%%%%%%%%%%%%%%%
\section{Introduction}
%%%%%%%%%%%%%%%%%%%%%%%%%%%%%%%%%%%%%%%%%%%%%%%%%%%%%%%%%%%%%%%%%%%%%%%%%%%%%%%%%%%%%%%%%%%%%%%%%%%%%%
The motility of bacteria near solid-liquid interfaces or on a solid surface plays a crucial role in fields such as biomedicine~\cite{Tuson2013}, and wastewater treatment~\cite{Liu2013,Ghosh2019}. Planktonic microorganisms, particularly flagellated bacteria, form biofilms through processes that include surface entrapment, migration, and colonization~\cite{Monds2009,Carniello2018,Krsmanovic2021}. The resulting biofilms are closely associated with various infectious diseases~\cite{Costerton1999,Donlan2001,Chang2018,Penesyan2021}. Flagellated bacteria propel themselves to swim in fluid environments through the thrust generated by rotating their helical flagella~\cite{Lauga2016,Liu2025B}. Recent experiments using a combination of three-dimensional holographic microscopy and optical tweezers to track an individual \emph{E. coli} have revealed that surface entrapment of bacteria can be divided into three sequential stages: approach, reorientation, and surface swimming~\cite{Bianchi2017,Bianchi2019}. The physical properties of the solid surface significantly influence the dynamic behavior of bacteria~\cite{Krsmanovic2021}. In particular, on no-slip solid surfaces, near-field hydrodynamic interactions (HIs) between bacteria and the surface are the primary mechanism for surface entrapment~\cite{Berke2008,Li2011,Wu2018,Zhang2021}. However, HIs alone are insufficient to fully explain the circular motion of bacteria on surfaces. The Derjaguin–Landau–Verwey–Overbeek (DLVO) forces prevent bacteria from contacting the surface~\cite{Sharma2003,Hong2012,Liu2025D}. In the phase plane, the bacterial motion exhibits a stable fixed point under the combination of near-field HIs and DLVO forces~\cite{Frymier1995,Li2008,Liu2025D}.

Bacteria undergoing Brownian motion near a no-slip solid surface exhibit various patterns of motility, including surface entrapment~\cite{Vissers2018,Berne2018}, circular motion~\cite{Li2008,Utada2014}, and escape from surfaces~\cite{Junot2022}. Both experiments and numerical simulations have demonstrated that the translational and rotational velocities, the radius of curvature of the circular motion, and the surface residence time of bacteria are mainly influenced by near-field HIs between bacteria and surfaces~\cite{Lauga2006,Junot2022}, Brownian motion~\cite{Li2008}, surface properties~\cite{Khalid2020,Zheng2021,Krsmanovic2021}, and bacterial morphology~\cite{Di2011,Tokarova2021}. The vertical translational velocity of bacteria is reduced as their distance to the surface decreases due to the near-field HIs~\cite{Lauga2006,Liu2025A}. The shear flow induced by a no-slip solid surface exerts a torque on the cell body that tends to orient the bacteria toward the surface~\cite{Petroff2015,Bianchi2017}. In contrast, the drag force acting on the flagella drives the bacteria to orient parallel to the surface~\cite{Sipos2015}. Consequently, bacteria with shorter flagella often orient perpendicularly to the surface ("nose down")~\cite{Petroff2015}, representing the surface-bound state where the flagellum points into the surface\cite{Das2019}. However, those with longer flagella tend to swim parallel to the surface, corresponding to the circular-swimming state~\cite{Petroff2018,Das2019,Liu2025D}. Moreover, experiments indicate that the surface suppresses bacterial tumbling, preventing the escape of bacteria from the solid surface~\cite{Molaei2014}. Random forces and torques are generally considered the primary mechanism that allows bacteria to escape surface entrapment~\cite{Schaar2015,Militaru2021}, and longer flagella have been reported to facilitate such escape~\cite{Sipos2015}. Experiments show that the surface residence time of bacteria that swim in the surface region can exceed one hundred seconds~\cite{Junot2022}. These observations highlight the important role of near-field HIs of bacteria in bacterial dynamics near solid surfaces. However, the detailed role of near-field HIs in bacterial entrapment and escape processes remains poorly understood, motivating the present study.

The goal of this article is to investigate how the near-field HIs between bacteria and a no-slip solid surface influence bacterial velocities, the radius of curvature of circular motion, and their surface residence time. In our numerical simulations, bacteria are represented by a chiral two-body model~\cite{Di2011,Dvoriashyna2021,Liu2025C}. The $6\times6$ resistance matrix of flagella is derived using resistive force theory (RFT)~\cite{Gray1955,Chwang1975,Johnson1979}, which includes flagellar morphology and chirality. The resistance matrix between the bacterial cell body and the surface is calculated using the method proposed by Dunstan \emph{et al.}, which accounts for near- and far-field HIs~\cite{Dunstan2012}.

This article is organized as follows. The numerical simulation model and methods are presented in Sec.~\ref{Sect2}, where we employ a chiral two-body model to represent the bacterium and to account for near-field HIs. Numerical simulations without Brownian motion and with Brownian motion are performed in Secs.~\ref{Sect3.1} and \ref{Sect3.2}, respectively.

%%%%%%%%%%%%%%%%%%%%%%%%%%%%%%%%%%%%%%%%%%%%%%%%%%%%%%%%%%%%%%%%%%%%%%%%%%%%%%%%%%%%%%%%%%%%%%%%%%%%%%
\section{Simulation Model}
\label{Sect2}
%%%%%%%%%%%%%%%%%%%%%%%%%%%%%%%%%%%%%%%%%%%%%%%%%%%%%%%%%%%%%%%%%%%%%%%%%%%%%%%%%%%%%%%%%%%%%%%%%%%%%%
\begin{figure}
\centering
\includegraphics[width=0.50\textwidth]{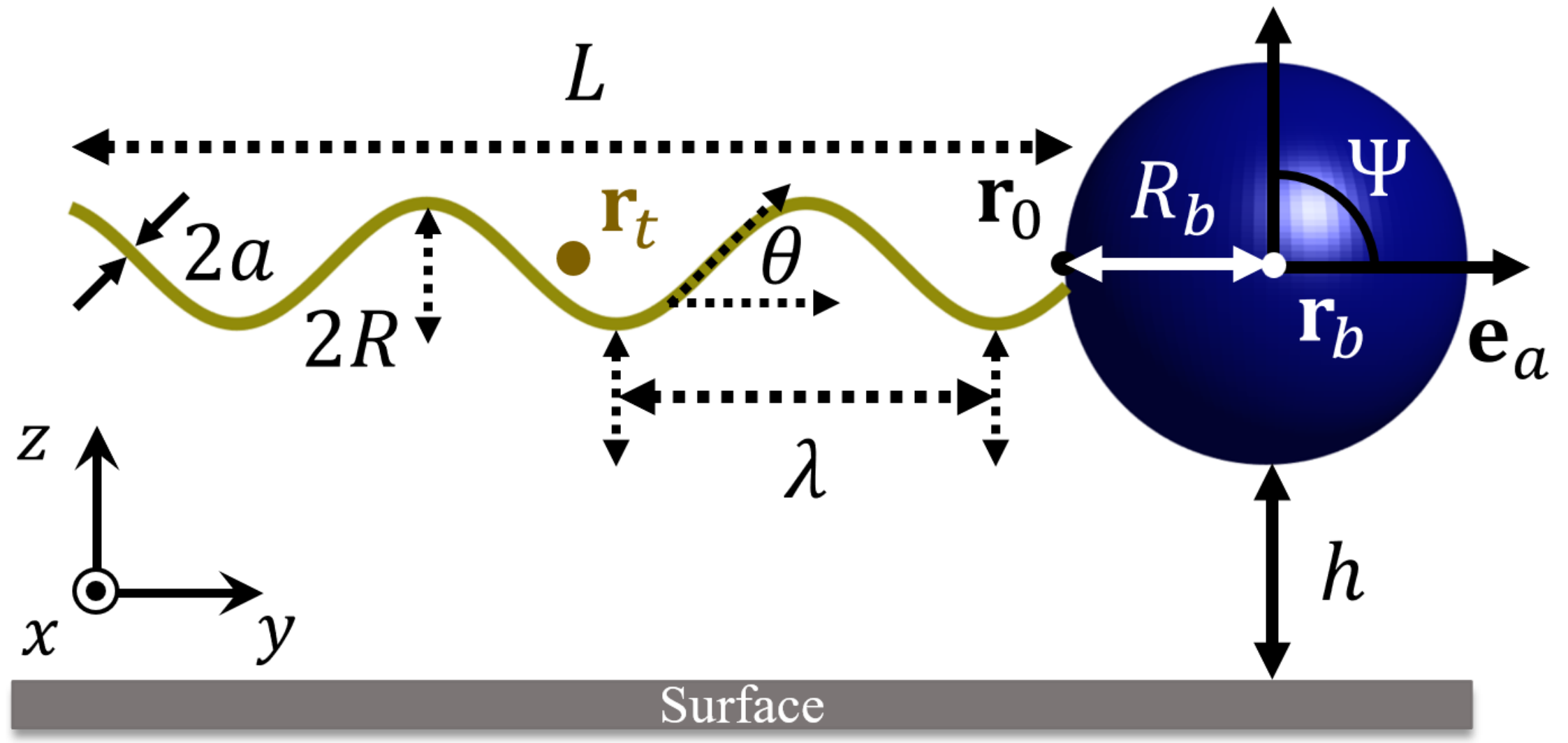}
\caption{Schematic diagram of a bacterium model with a rigid helical flagellum and a spherical cell body swimming near a no-slip solid surface. $\mathbf{r}_{b}$ and $\mathbf{r}_{t}$ are the center positions of the cell body and flagellum, respectively. The closest distance between the cell body and solid surface is $h$, and the inclination angle is $\Psi$. The surface is located on the $xy$-plane.}
\label{fig:fig1}
\end{figure}

A single flagellated bacterium is modeled as a single left-handed helical flagellum attached to a spherical cell body, which is used as a simplified model for \emph{E. coli}~\cite{Lauga2006,Li2008,Sipos2015,Petroff2015}. The closest distance between the cell body and the no-slip solid surface is $h$, as shown in Fig.~\ref{fig:fig1}. The surface is located on the $xy$-plane, and the connection point between the cell body and the flagellar axis is denoted as $\mathbf{r}_{0}$. The helical flagellum is modeled as a rigid helix with filament radius $a$, helix radius $R$, pitch $\lambda$, pitch angle $\theta$, axial length $L$, and contour length $\Lambda=L/\cos\theta$, where $\tan\theta=2\pi R/\lambda$. The centerline positions of the left-handed helical flagellum are given by~\cite{Rodenborn2013,Esparza2021,Liu2025B}:
\begin{equation}
\begin{split}
\mathbf{r}(l)=l\cos\theta\mathbf{D}_{1}+R\sin\phi\mathbf{D}_{2}+R\cos\phi\mathbf{D}_{3}+\mathbf{r}_{0}.
\label{eq:refname01}
\end{split}
\end{equation}
where $l\in[0,\Lambda]$ is the arc length along the centerline, $k=2\pi/\lambda$ is the wave number, and $\phi=kl\cos\theta$ is the phase. The time-dependent material frame at the connection point $\mathbf{r}_{0}$ is denoted as $\{\mathbf{D}_{1},\mathbf{D}_{2},\mathbf{D}_{3}\}$. This is a set of orthonormal vectors, with $\mathbf{D}_{1}$ being the unit vector along the flagellar axis. The radius of the spherical cell body is $R_{b}$. The bacterium can be represented by a chiral two-body model, with the centers of the cell body and the flagellum denoted as $\mathbf{r}_{b}$ and $\mathbf{r}_{t}$, respectively. This model can be effectively used to simulate the Brownian motion of \emph{E. coli}~\cite{Liu2025C}. The distance between the centers of the cell body and the flagellum is $d=L/2+R_{b}$. The points $\mathbf{r}_{0}$ and $\mathbf{r}_{b}$ satisfy the relationship: $\mathbf{r}_{0}=\mathbf{r}_{b}-R_{b}\cdot\mathbf{D}_{1}$.

\begin{table}[h]
\small
\caption{Geometric parameters and corresponding values of a bacterium system~\cite{Lauga2006,Li2008,Liu2025D}}.
\label{tab:table01}
\begin{tabular*}{0.48\textwidth}{@{\extracolsep{\fill}}lll}
\hline
\textbf{Symbol} & \textbf{Parameters} & \textbf{Values}\\
\hline
$\mu$ & Dynamic viscosity & $1.0$~\si{\mu g/(\mu m\cdot s)}\\
$R_{b}$ & Radius of cell body & $1.0$~\si{\mu m}\\
$a$ & Filament radius & $0.01$~\si{\mu m}\\
$R$ & Helix radius & $0.25$~\si{\mu m}\\
$\theta$ & Pitch angle & $\pi/5$\\
$\Lambda$ & Contour length & $7.5$~\si{\mu m}\\
$f$ & Rotation frequency of motor & $100$~\si{Hz}\\
$H$ & Hamaker constant & $10^{-21}$~\si{J}\\
$\zeta_{1}$ & Zeta potential of cell body & $-20.0$~\si{mV}\\
$\zeta_{2}$ & Zeta potential of surface & $-20.0$~\si{mV}\\
\hline
\end{tabular*}
\end{table}

HIs between the cell body and flagellum, and between the flagellum and the surface, are neglected. HIs between the cell body and the flagellum contribute only a small error to the final results, because the presence of a nearby surface leads to spatially localized flow fields that decay as rapidly as a Stokeslet-dipole~\cite{Lauga2006}. HIs between the flagellum and surface become significant only at separations less than the filament radius ($a=10$~\si{nm}), a separation that only a small part of flagella can achieve.~\cite{Li2008}. The rigid helical flagellum is represented by a chiral body model, with its $6\times6$ resistance matrix obtained by integrating the flagellar centerline using RFT and expressed as~\cite{Liu2025C}
\begin{equation}
\begin{split}
\mathcal{R}_{t}=
\left(\begin{matrix} 
A_{f} & B_{f}^{T} \\
B_{f} & C_{f}
\end{matrix}\right).
\label{eq:refname02}
\end{split}
\end{equation}
where the three submatrices are given by~\cite{Di2011,Dvoriashyna2021,Liu2025C}:
\begin{equation}
\begin{split}
&A_{f}=X_{\parallel}^{A}\mathbf{e}_{a}\otimes\mathbf{e}_{a}+X_{\perp}^{A}\left(\mathbb{I}-\mathbf{e}_{a}\otimes\mathbf{e}_{a}\right),\\
&B_{f}=X_{\parallel}^{B}\mathbf{e}_{a}\otimes\mathbf{e}_{a}+X_{\perp}^{B}\left(\mathbb{I}-\mathbf{e}_{a}\otimes\mathbf{e}_{a}\right),\\
&C_{f}=X_{\parallel}^{C}\mathbf{e}_{a}\otimes\mathbf{e}_{a}+X_{\perp}^{C}\left(\mathbb{I}-\mathbf{e}_{a}\otimes\mathbf{e}_{a}\right).
\label{eq:refname03}
\end{split}
\end{equation}
where $\mathbf{e}_{a}=\mathbf{D}_{1}$ is the unit vector that indicates the direction of the flagellar axis. The specific expressions of these matrix elements are detailed in Appendix A. The resistance matrix for a spherical cell body near a solid surface can be written as~\cite{Dunstan2012,Liu2025D}
\begin{equation}
\begin{split}
\mathcal{R}_{b}=\left(\begin{matrix} 
A_{b} & B_{b}^{T} \\
B_{b} & C_{b}
\end{matrix}\right)
=\left(\begin{matrix} 
Y_{\parallel}^{A} & 0 & 0 & 0 & Y^{B} & 0 \\
0 & Y_{\parallel}^{A} & 0 & -Y^{B} & 0 & 0 \\
0 & 0 & Y_{\perp}^{A} & 0 & 0 & 0 \\
0 & -Y^{B} & 0 & Y_{\parallel}^{C} & 0 & 0 \\
Y^{B} & 0 & 0 & 0 & Y_{\parallel}^{C} & 0\\
0 & 0 & 0 & 0 & 0 & Y_{\perp}^{C} 
\end{matrix}\right).
\label{eq:refname04}
\end{split}
\end{equation}
Detailed expressions of these matrix elements are provided in Appendix B.

\begin{figure}
\centering
\includegraphics[width=0.50\textwidth]{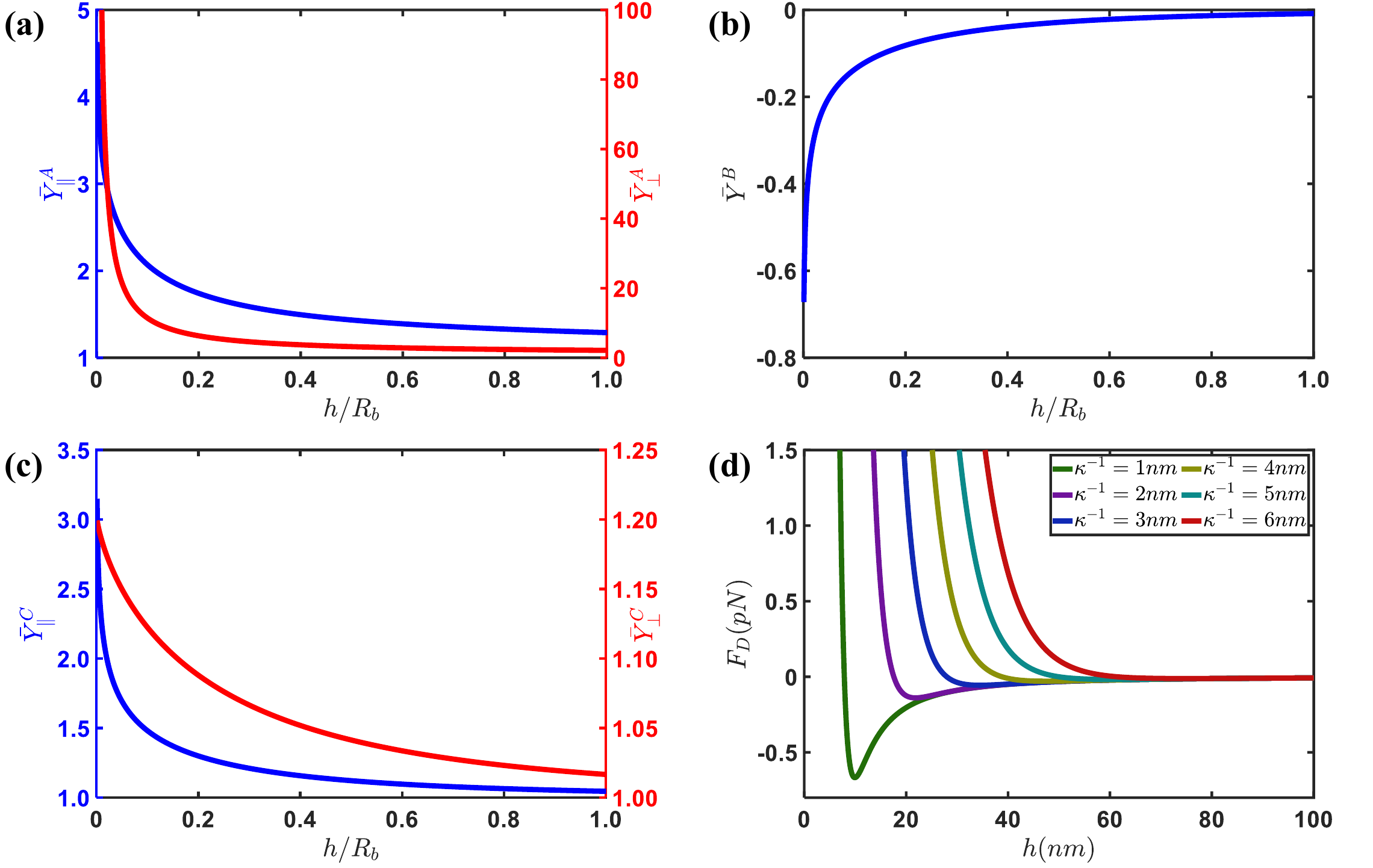}
\caption{The scalar functions (a) $\bar{Y}_{\parallel}^{A}$, $\bar{Y}_{\perp}^{A}$, (b) $\bar{Y}^{B}$, (c) $\bar{Y}_{\parallel}^{C}$ and $\bar{Y}_{\perp}^{C}$ of the resistance matrix as a function of reduced height $h/R_{b}$. (d) DLVO forces of a sphere interacting with a solid surface as a function of the closest distance $h$ for different Debye lengths.}
\label{fig:fig2}
\end{figure}

The scalar functions $\bar{Y}_{\parallel}^{A}$, $\bar{Y}_{\perp}^{A}$, $\bar{Y}^{B}$, $\bar{Y}_{\parallel}^{C}$, and $\bar{Y}_{\perp}^{C}$ of the resistance matrix of the cell body are plotted in Fig.~\ref{fig:fig2}. It is evident that these functions are strongly dependent on the reduced height $h/R_{b}$. In particular, the function $Y_{\perp}^{A}$ at $h/R_{b}=0.1$ is approximately $10$ times larger than its value in the bulk fluid ($h/R_{b}\to\infty$), and diverges as $h/R_{b}$ approaches $0$. The bacterial diffusion coefficient, $D=k_{B}T/\mathcal{R}$, where $k_{B}$ is the Boltzmann constant, $T$ is the absolute temperature and $\mathcal{R}$ is the resistance matrix. As bacteria approach the surface, their resistance increases, resulting in a decrease in their diffusion coefficient and effectively suppressing their motility, as shown in Appendix C. The functions $Y_{\parallel}^{A}$, $Y_{\parallel}^{C}$, and $Y_{\perp}^{C}$ also increase as $h$ decreases, indicating that the bacterial velocities decrease closer to the surface. The function $Y^{B}$ characterizes the hydrodynamic coupling between rotational velocity and force, which can result in "nose down" configurations~\cite{Das2015,Sipos2015,Petroff2015,Bianchi2019,Liu2025A}.

In addition to the HIs, the DLVO force is also present between the cell body and the surface. The DLVO force comprises the sum of van der Waals (vdW) force and electrostatic (ele) force~\cite{Sharma2003,Hong2012,Liu2025D}:
\begin{equation}
\begin{split}
&F_{D}(h)=F_{vdW}(h)+F_{ele}(h).
\label{eq:refname05}
\end{split}
\end{equation}
where $h$ is the closest distance between the cell body and the surface. The expressions for the van der Waals (vdW) force and the electrostatic (ele) force are given by~\cite{Sharma2003,Hong2012,Liu2025D}:
\begin{equation}
\begin{split}
&F_{vdW}(h)=-\frac{H}{3}\frac{R_{b}^{3}}{h^{2}(h+2R_{b})^{2}},\\
&F_{ele}(h)=2\pi\varepsilon\kappa R_{b}\frac{2\zeta_{1}\zeta_{2}e^{\kappa h}-(\zeta_{1}^{2}+\zeta_{2}^{2})}{e^{2\kappa h}-1}.
\label{eq:refname06}
\end{split}
\end{equation}
where $\varepsilon=6.933\times10^{-10}$~\si{C^{2}/(N\cdot m^{2})} is the permittivity of water at room temperature $T=298$~\si{K}, $H$ is the Hamaker constant, and $\zeta_{1}$ and $\zeta_{2}$ are the zeta potentials of the cell body and the solid surface, respectively. The Debye length, $\kappa^{-1}$, is given by $\kappa^{-1}=0.304/\sqrt{I}$\si{nm}, where $I$ is the ionic strength (in \si{mol/L})~\cite{Liu2025D}.

In the present paper, we set $H=10^{-21}$~\si{J} and the zeta potentials as $\zeta_{1}=\zeta_{2}=-20$~\si{mV}~\cite{Vigeant2002,Li2008}. Fig.~\ref{fig:fig2}(d) illustrates the DLVO forces between a sphere of radius $R_{b}=1$~\si{\mu m} and a solid surface for different Debye lengths. As the sphere approaches the surface, the DLVO forces increase rapidly with decreasing separation distance $h$, which can effectively prevent bacteria from contacting the surface during the Brownian motion. Bacteria reach a stable fixed point $\{h^{*},\Psi^{*}\}$ in the phase plane $\{h,\Psi\}$ under the combination of near-field HIs and DLVO forces~\cite{Liu2025D}. HIs are long-range forces, yet their influence becomes particularly significant at separations less than the particle radius, while DLVO forces are short-range forces, with an effective distance set at $100$~\si{nm}. The angle of inclination $\Psi$, defined as the angle between the flagellar axis and the $z$-axis, characterizes the orientation of the bacteria. Specifically, bacteria tend to exhibit circular motion on the surface at this stable fixed point~\cite{Frymier1995}.

The instantaneous translational and rotational velocities of the center of the cell body are $\mathbf{U}_{b}=\{U_{x},U_{y},U_{z}\}$ and $\mathbf{W}_{b}=\{W_{x},W_{y},W_{z}\}$, respectively, while those of the flagellar center are
\begin{equation}
\begin{split}
&\mathbf{W}_{t}=\mathbf{W}_{b}+\mathbf{W}_{0},\\
&\mathbf{U}_{t}=\mathbf{U}_{b}+\mathbf{W}_{t}\times(\mathbf{r}_{t}-\mathbf{r}_{b}).
\label{eq:refname07}
\end{split}
\end{equation}
where $\mathbf{W}_{0}=2\pi f\mathbf{e}_{m}$ is the rotational velocity of the motor, and $f$ is its rotation frequency. The rotation direction of the motor is along the flagellar axis in the present paper, that is, $\mathbf{e}_{m}=-\mathbf{D}_{1}$. The geometric parameters and the corresponding values of the bacterium system are presented in Table~\ref{tab:table01}.

The force and torque balance equations governing the motion of a bacterium swimming near a solid surface are given by:
\begin{equation}
\begin{split}
&\mathbf{F}_{b}^{h}+\mathbf{F}_{t}^{h}-\mathbf{F}_{D}-\mathbf{F}_{b}^{B}-\mathbf{F}_{t}^{B}=0,\\
&\mathbf{T}_{b}^{h}+\mathbf{T}_{t}^{h}-\mathbf{T}_{b}^{B}-\mathbf{T}_{t}^{B}+\left(\mathbf{r}_{t}-\mathbf{r}_{b}\right)\times\left(\mathbf{F}_{t}^{h}-\mathbf{F}_{t}^{B}\right)=0.
\label{eq:refname08}
\end{split}
\end{equation}
where $\mathbf{F}_{b}^{h}$, $\mathbf{T}_{b}^{h}$, $\mathbf{F}_{t}^{h}$, and $\mathbf{T}_{t}^{h}$ are the forces and torques exerted on the fluid by the cell body and the flagellum, respectively. $\mathbf{F}_{D}$ is the DLVO force between the cell body and the solid surface. $\mathbf{F}_{b}^{B}$, $\mathbf{T}_{b}^{B}$, $\mathbf{F}_{t}^{B}$, and $\mathbf{T}_{t}^{B}$ represent the random forces and torques on the cell body and flagellum, respectively. The expressions for these random forces are as follows~\cite{Li2008,Liu2025C}:
\begin{equation}
\begin{split}
&\mathbf{F}_{b}^{B}=\sqrt{\frac{2k_{B}T}{\Delta t}}\sqrt{(\mathcal{R}_{b})_{ii}}\xi_{1},\quad i=1,2,3,\\
&\mathbf{F}_{t}^{B}=\sqrt{\frac{2k_{B}T}{\Delta t}}\sqrt{(\mathcal{R}_{t})_{ii}}\xi_{2},\quad i=1,2,3.
\label{eq:refname09}
\end{split}
\end{equation}
and the random torques are~\cite{Li2008,Liu2025C}
\begin{equation}
\begin{split}
&\mathbf{T}_{b}^{B}=\sqrt{\frac{2k_{B}T}{\Delta t}}\sqrt{(\mathcal{R}_{b})_{(i+3)(i+3)}}\xi_{3},\quad i=1,2,3,\\
&\mathbf{T}_{t}^{B}=\sqrt{\frac{2k_{B}T}{\Delta t}}\sqrt{(\mathcal{R}_{t})_{(i+3)(i+3)}}\xi_{4},\quad i=1,2,3.
\label{eq:refname10}
\end{split}
\end{equation}
where $\xi_{1}$, $\xi_{2}$, $\xi_{3}$ and $\xi_{4}$ are independent Gaussian random variables with zero mean and unit variance. The time step $\Delta t=10^{-4}$~\si{s} in our simulations.

When a bacterium is submerged in a viscous fluid, its motion is governed by the following equation~\cite{Lauga2006,Lauga2009,Liu2025D}:
\begin{equation}
\left(\begin{matrix}\mathcal{R}_{b} & \mathbf{0}\\ \mathbf{0} & \mathcal{R}_{t}\end{matrix}\right) \left(\begin{matrix}\mathbf{U}_{b}\\ \mathbf{W}_{b}\\ \mathbf{U}_{t}\\ \mathbf{W}_{t} \end{matrix}\right)=\left(\begin{matrix}\mathbf{F}_{b}^{h}\\ \mathbf{T}_{b}^{h} \\ \mathbf{F}_{t}^{h}\\ \mathbf{T}_{t}^{h} \end{matrix}\right).
\label{eq:refname11}
\end{equation}
where $\mathbf{0}$ is a $6\times6$ zero matrix. The center of the cell body is updated by~\cite{Liu2025B}:
\begin{equation}
\begin{split}
&\mathbf{r}_{b}(t+\Delta t)=\mathbf{r}_{b}(t)+\mathbf{U}_{b}(t)\Delta t.
\label{eq:refname12}
\end{split}
\end{equation}
and the material frame $\{\mathbf{D}_{1},\mathbf{D}_{2},\mathbf{D}_{3}\}$ is updated as follows~\cite{Liu2025B}:
\begin{equation}
\begin{split}
&\{\mathbf{D}_{1}(t+\Delta t),\mathbf{D}_{2}(t+\Delta t),\mathbf{D}_{3}(t+\Delta t)\}\\
&=\mathbf{R}(\mathbf{e}_{b},\theta_{b})\cdot\mathbf{R}(\mathbf{e}_{0},\theta_{0})\cdot\{\mathbf{D}_{1}(t),\mathbf{D}_{2}(t),\mathbf{D}_{3}(t)\}.
\label{eq:refname13}
\end{split}
\end{equation}
where $\mathbf{e}_{b}=\mathbf{W}_{b}(t)/|\mathbf{W}_{b}(t)|$, $\theta_{b}=|\mathbf{W}_{b}(t)|\Delta t$, $\mathbf{e}_{0}=\mathbf{W}_{0}(t)/|\mathbf{W}_{0}(t)|$ and $\theta_{0}=|\mathbf{W}_{0}(t)|\Delta t$. The rotation matrix $\mathbf{R}(\mathbf{e},\theta)$ is Rodrigues' rotation matrix, where $\mathbf{e}$ indicates the direction of the rotation axis and $\theta$ is the rotation angle. Bacterial trajectories can be effectively simulated using this algorithm.

%%%%%%%%%%%%%%%%%%%%%%%%%%%%%%%%%%%%%%%%%%%%%%%%%%%%%%%%%%%%%%%%%%%%%%%%%%%%%%%%%%%%%%%%%%%%%%%%%%%%%%
\section{Results}
\label{Sect3}
\subsection{Surface Swimming Dependent on Near-Field HIs}
\label{Sect3.1}
%%%%%%%%%%%%%%%%%%%%%%%%%%%%%%%%%%%%%%%%%%%%%%%%%%%%%%%%%%%%%%%%%%%%%%%%%%%%%%%%%%%%%%%%%%%%%%%%%%%%%%
We first qualitatively and mechanistically examine the effect of near-field HIs on bacterial motility near a no-slip solid surface, excluding Brownian motion. When considering only near-field HIs, the translational and rotational velocities of the cell body as it gradually approaches the surface perpendicularly are:
\begin{equation}
\begin{split}
&U_{z}=\frac{Y_{\perp}^{C}X_{\parallel}^{B}}{(X_{\parallel}^{B})^{2}-(X_{\parallel}^{A}+Y_{\perp}^{A})(X_{\parallel}^{C}+Y_{\perp}^{C})}W_{0},\\
&W_{z}=\frac{X_{\parallel}^{A}X_{\parallel}^{C}-(X_{\parallel}^{B})^{2}+Y_{\perp}^{A}X_{\parallel}^{C}}{(X_{\parallel}^{B})^{2}-(X_{\parallel}^{A}+Y_{\perp}^{A})(X_{\parallel}^{C}+Y_{\perp}^{C})}W_{0}.
\label{eq:refname14}
\end{split}
\end{equation}

\begin{figure}
\centering
\includegraphics[width=0.50\textwidth]{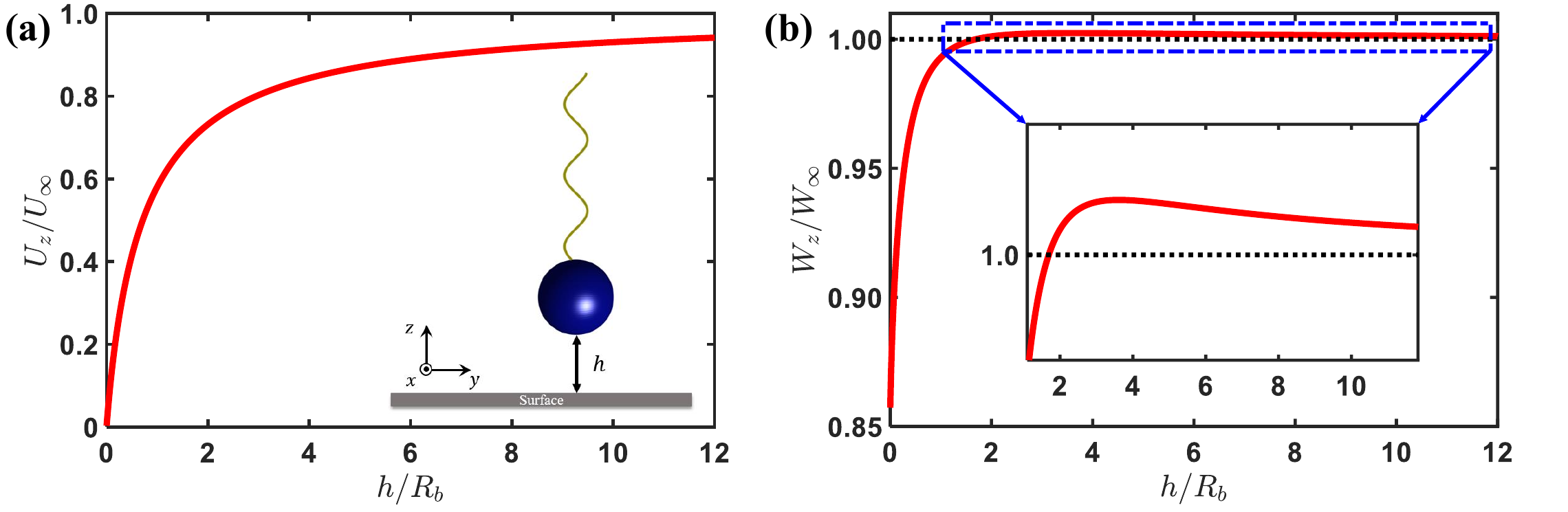}
\caption{The normalized velocities of a bacterium approaching the surface perpendicularly. (a) The vertical translational velocity $U_{z}$ as a function of reduced height $h/R_{b}$ normalized by the bulk velocity $U_{\infty}$. The inset is a schematic. (b) The vertical rotational velocity $W_{z}$ as a function of reduced height $h/R_{b}$ normalized by the bulk velocity $W_{\infty}$.}
\label{fig:fig3}
\end{figure}

As shown in Fig.~\ref{fig:fig3}, we plot the normalized vertical translational velocity of the cell body, $U_{z}/U_{\infty}$, and rotational velocities, $W_{z}/W_{\infty}$, as a function of the reduced height $h/R_{b}$. The translational velocity of bacteria in the bulk fluid is approximately $U_{\infty}=14.5$~\si{\mu m/s} for the parameters given in Table~\ref{tab:table01}, which is consistent with the experimental measurements~\cite{Molaei2014}. In Figs.~\ref{fig:fig2}(a) and (c), the resistance function $Y_{\perp}^{A}$ increases significantly faster than $Y_{\perp}^{C}$ as the separation distance $h$ decreases. Consequently, the first equation of Eq.~\ref{eq:refname14} implies that the vertical translational velocity $U_{z}$ monotonically decreases as the separation distance $h$ decreases, as depicted in Fig.~\ref{fig:fig3}(a). The function $Y_{\perp}^{A}$ diverges to infinity as $h$ approaches zero, resulting in the vertical translational velocity of the cell body approaching zero. However, the DLVO force prevents the bacteria from contacting the surface. There is a stable separation distance between the cell body and the surface, which typically ranges from a few nanometers to over a hundred nanometers~\cite{Petroff2018,Liu2025D}.

Similarly, Fig.~\ref{fig:fig3}(b) shows the normalized vertical rotational velocity of the cell body, $W_{z}/W_{\infty}$, as a function of the reduced height $h/R_{b}$. This rotational velocity is also determined by the coupling coefficients $Y_{\perp}^{A}$ and $Y_{\perp}^{C}$, as shown in the second equation of Eq.~\ref{eq:refname14}. In the absence of a surface, the bacterial rotational velocity is approximately $W_{\infty}=18.3$~\si{rad/s}, according to the parameters given in Table~\ref{tab:table01}. Because RFT neglects the hydrodynamic interaction between the flagellum and the cell body, the rotational velocity of the cell body is lower than the experimental measurement~\cite{Liu2025C}. However, for qualitatively investigating the velocity change of bacteria approaching the surface, this discrepancy is acceptable. Observing the second equation of Eq.~\ref{eq:refname14}, both the numerator and the denominator inherently involve the resistance term $Y_{\perp}^{A}$. As the separation distance $h$ decreases, the vertical rotational velocity of the cell body, $W_{z}$, is jointly influenced by $Y_{\perp}^{A}$ and $Y_{\perp}^{C}$. These functions represent the resistance components for the translational and rotational velocities of the cell body, respectively. The net rotational velocity arises from the competition between these two resistance functions. Consequently, as depicted in Fig.~\ref{fig:fig3}(b), the vertical rotational velocity exhibits a slight increase initially before decreasing.

\begin{figure}
\centering
\includegraphics[width=0.50\textwidth]{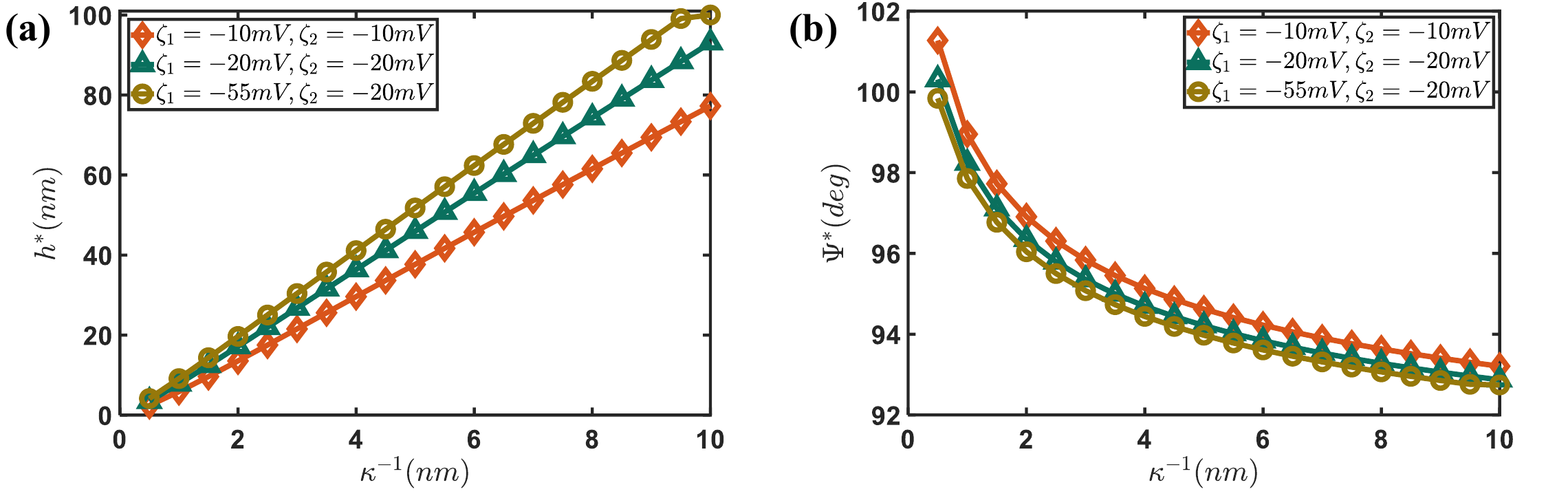}
\caption{(a) Stable heights $h^{*}$ and (b) stable inclination angles $\Psi^{*}$ as a function of Debye length for different values of zeta potentials $\zeta$.}
\label{fig:fig4}
\end{figure}

During the bacterial surface entrapment process, only hydrodynamic and DLVO interactions are considered. Bacterial surface entrapment is typically divided into three sequential stages: approach, reorientation, and surface swimming~\cite{Bianchi2017,Bianchi2019,Liu2025D}. Near a no-slip solid surface, a stable fixed point $\{h^{*},\Psi^{*}\}$ exists within the phase plane $\{h,\Psi\}$~\cite{Shum2010,Shum2015}, where the bacterium performs stable circular motions on the surface. Fig.~\ref{fig:fig4} illustrates the dependence of stable heights ($h^{*}$) and stable inclination angles ($\Psi^{*}$) on the Debye length, with curves shown for varying zeta potentials. These stable states are mainly determined by the bacterial morphology and the Debye length~\cite{Liu2025D}.

\begin{figure}
\centering
\includegraphics[width=0.50\textwidth]{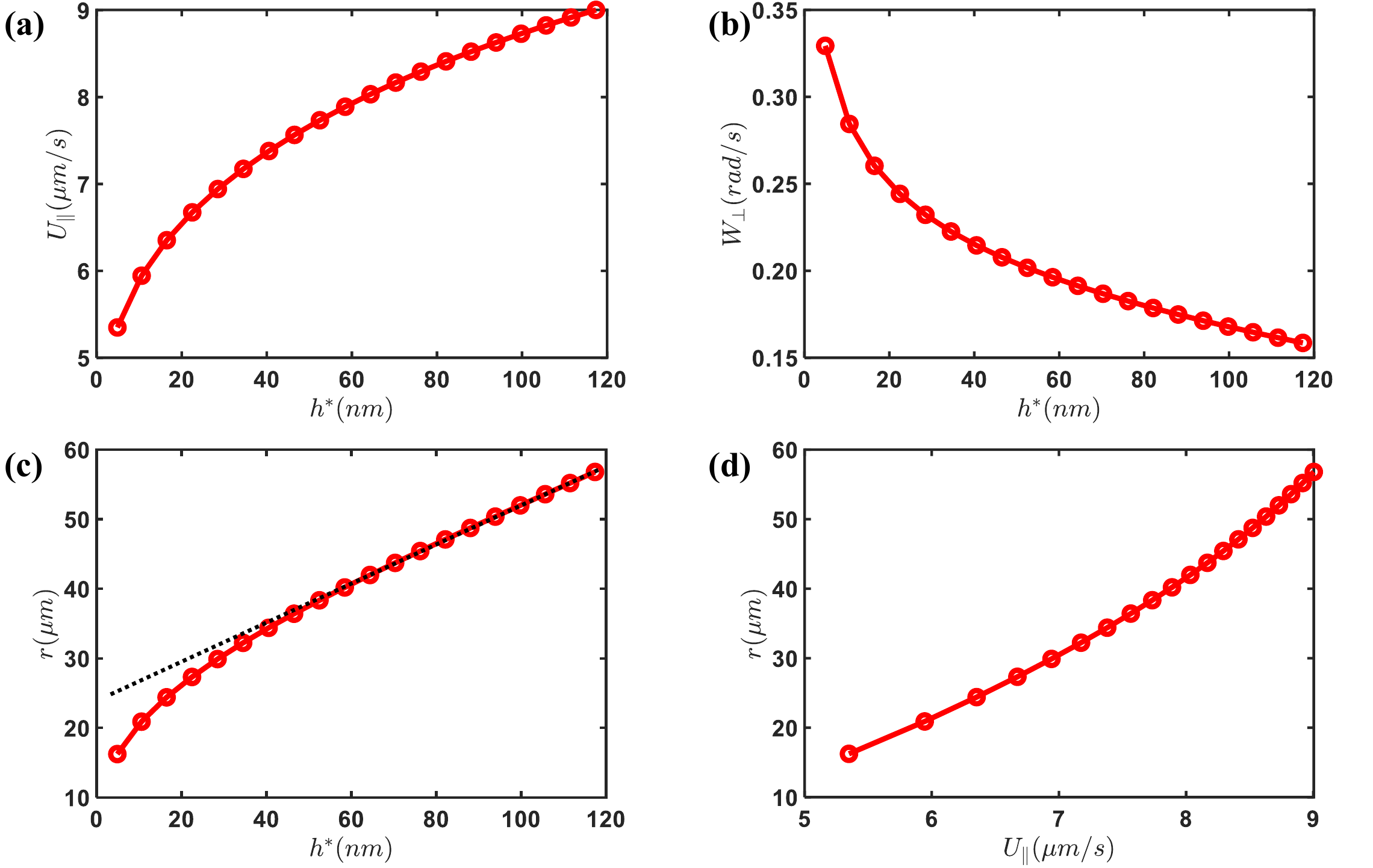}
\caption{Bacterial velocities and radius of curvature of their circular motion in the surface swimming stage. (a) Horizontal translational velocity, $U_{\parallel}$, as a function of stable height $h^{*}$. (b) Vertical rotational velocity, $W_{\perp}$, as a function of stable height $h^{*}$. (c) Radius of curvature, $r$, as a function of stable height $h^{*}$. (d) Radius of curvature, $r$, as a function of horizontal translational velocity $U_{\parallel}$.}
\label{fig:fig5}
\end{figure}

Experimental observations have revealed that the radius of curvature is dependent on the bacterial swimming speed~\cite{Liu2024}. Under conditions of fixed bacterial morphology and motor rotation rate, the horizontal translational velocity and vertical rotational velocity of bacteria are solely determined by the stable fixed point $\{h^{*},\Psi^{*}\}$. Bacteria exhibit circular motion near a solid surface during the surface swimming stage~\cite{Lauga2006}. The radius of curvature, $r$, of this circular motion is given by~\cite{Liu2025D}
\begin{equation}
\begin{split}
&r=\frac{U_{\parallel}}{W_{\perp}}=\frac{\sqrt{U_{x}^{2}+U_{y}^{2}}}{|W_{z}-\left(W_{x}\cos\phi+W_{y}\sin\phi\right)\cot\Psi^{*}|}.
\label{eq:refname15}
\end{split}
\end{equation}
where $\Psi^{*}$ is the stable inclination angle of bacteria, $\phi$ is the azimuth angle of the rotational velocity in spherical coordinates. The translational velocity of bacteria parallel to the surface is $U_{\parallel}=\sqrt{U_{x}^{2}+U_{y}^{2}}$, and the rotational velocity of bacteria perpendicular to the surface (along the $z$-axis) is $W_{\perp}=|W_{z}-\left(W_{x}\cos\phi+W_{y}\sin\phi\right)\cot\Psi^{*}|$. The terms $W_{x}$, $W_{y}$, $W_{z}$ are the components of the rotational velocity of the cell body in Cartesian coordinates.

Observations from Eqs.~\ref{eq:refnameB01}-\ref{eq:refnameB05} reveal that individual resistance functions of the cell body exhibit different dependencies on the separation distance $h$. Figs.~\ref{fig:fig5}(a)-(b) show that the horizontal translational velocity $U_{\parallel}$ increases with stable height $h^{*}$, while the vertical rotational velocity $W_{\perp}$ decreases as $h^{*}$ increases. As shown in Fig.~\ref{fig:fig5}(c), the radius of curvature of the bacterial circular motion increases monotonically with stable height. For stable heights that exceed $40$~\si{nm}, the radius of curvature exhibits a linear relationship with $h^{*}$~\cite{Liu2024}. As shown in Fig.~\ref{fig:fig5}(d), the radius of curvature also increases monotonically with the translational velocity $U_{\parallel}$~\cite{Liu2024}. These observed relationships are a consequence of near-field HIs, which are the physical mechanisms for surface entrapment in swimming bacteria. These near-field HIs, along with DLVO forces, collectively determine the stable height $h^{*}$ and the inclination angle $\Psi^{*}$, which subsequently influence the translational velocity, rotational velocity, and radius of curvature of the circular motion. However, the effective range of DLVO forces is significantly shorter, often by several orders of magnitude, than that of near-field HIs. Consequently, the bacterial dynamic behavior near the surface is dominated by near-field HIs.

%%%%%%%%%%%%%%%%%%%%%%%%%%%%%%%%%%%%%%%%%%%%%%%%%%%%%%%%%%%%%%%%%%%%%%%%%%%%%%%%%%%%%%%%%%%%%%%%%%%%%%
\subsection{Near-Field HIs Prolong Surface Residence Time}
\label{Sect3.2}
%%%%%%%%%%%%%%%%%%%%%%%%%%%%%%%%%%%%%%%%%%%%%%%%%%%%%%%%%%%%%%%%%%%%%%%%%%%%%%%%%%%%%%%%%%%%%%%%%%%%%%
The resistance matrix (Eq.~\ref{eq:refname04}) for the cell body near a no-slip solid surface comprises three submatrices: $A_{b}$, which describes the coupling between force and translational velocity; $C_{b}$, which describes the coupling between torque and rotational velocity; and $B_{b}$, which describes the coupling between torque and translational velocity. The elements of the submatrices $A_{b}$ and $C_{b}$ reveal that as the bacterium approaches the surface, the hydrodynamic resistance to its translational and rotational motions increases. This consequently leads to a reduction in both translational and rotational velocities, implying increasingly restricted motion for the bacterium near the surface. In contrast, the submatrix $B_{b}$ reveals that shear flow induced by the no-slip solid surface generates a rolling motion of the cell body, which in turn promotes the bacteria to tend to orient towards the surface~\cite{Das2015,Liu2025A}.

\begin{figure}
\centering
\includegraphics[width=0.40\textwidth]{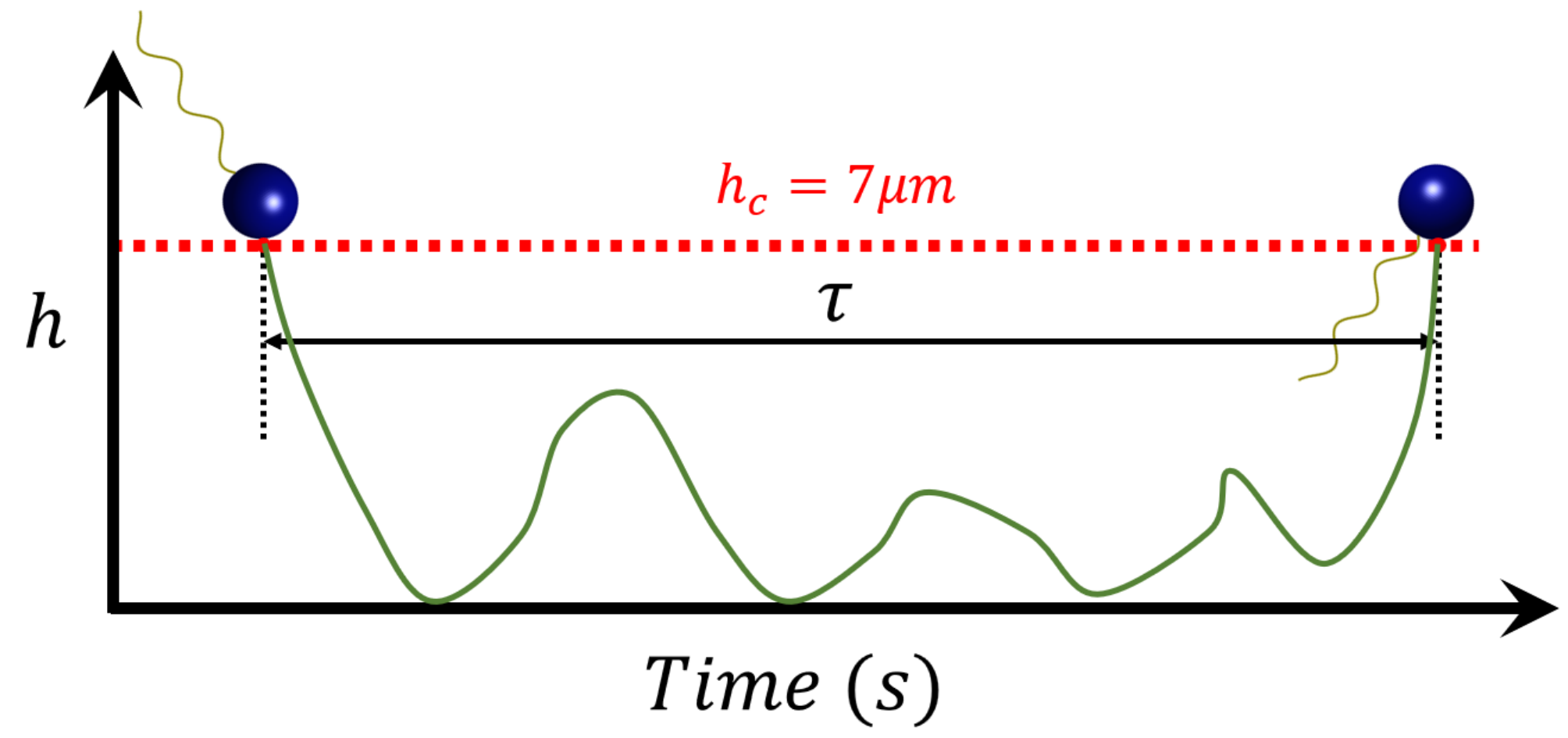}
\caption{Sketch diagram of a bacterial trajectory, the bacterium enters and leaves the surface at critical height $h_{c}=7$~\si{\mu m}. The surface residence time of bacteria is defined as $\tau$.}
\label{fig:fig6}
\end{figure}

In the absence of Brownian motion, bacteria are effectively trapped on the surface under the combination of near-field HIs and DLVO forces, exhibiting stable motion on the surface. However, the presence of Brownian motion introduces translational and rotational diffusion, allowing bacteria to eventually escape from the surface~\cite{Li2008,Junot2022}. However, as bacteria approach the surface, near-field HIs suppress both their translational and rotational diffusion (as shown in Appendix C), potentially prolonging their surface residence time. Furthermore, the non-zero submatrix $B_{b}$ promotes the orientation of bacteria towards the surface, further prolonging their surface residence time.

\begin{figure}
\centering
\includegraphics[width=0.50\textwidth]{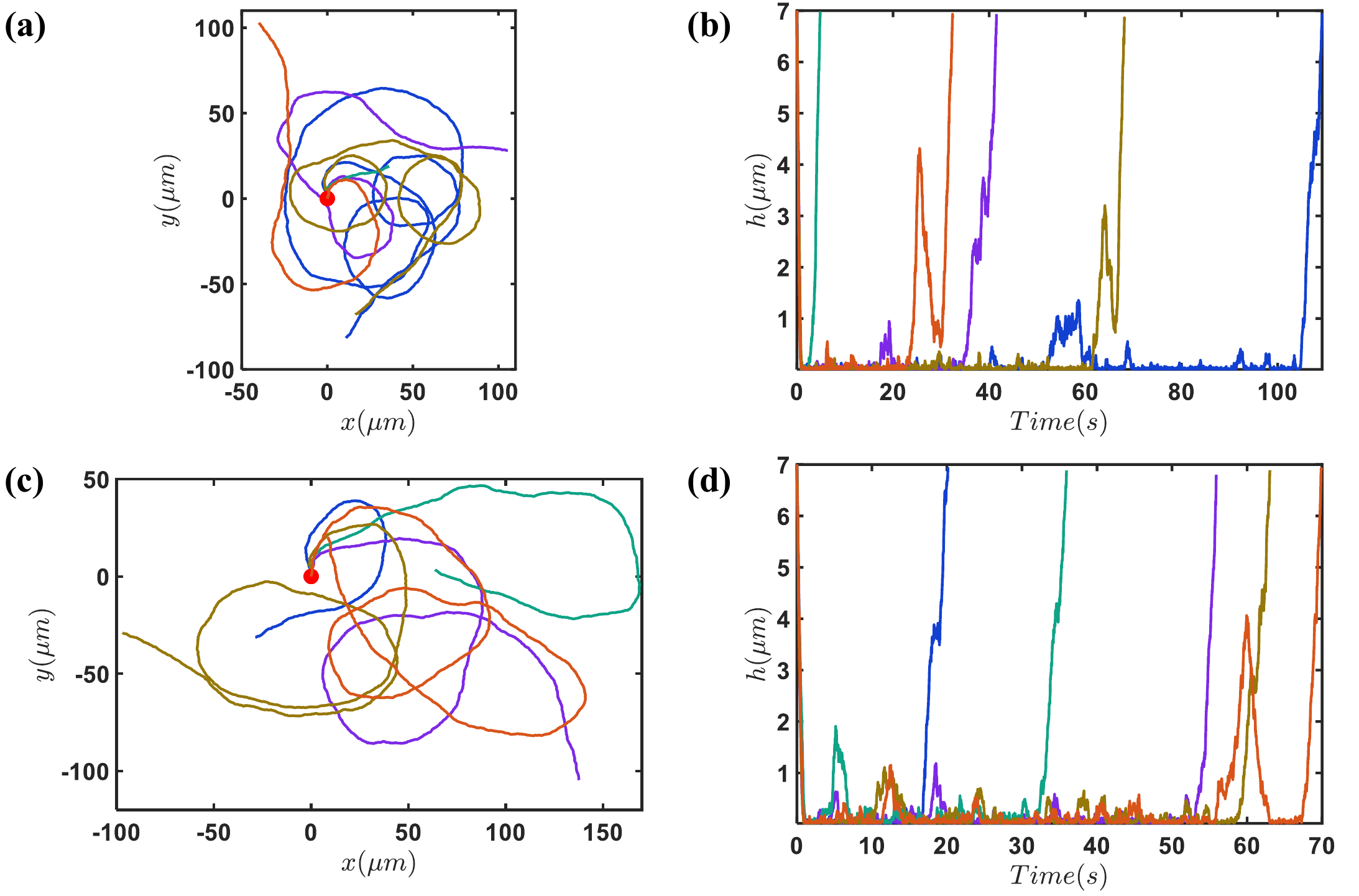}
\caption{Simulations conducted for Debye lengths: $\kappa^{-1}=2$~\si{nm} and $\kappa^{-1}=4$~\si{nm}. (a) Projection of five simulated bacterial trajectories onto the $xy$-plane for $\kappa^{-1}=2$~\si{nm}. (b) The corresponding heights $h$ as a function of time for these five trajectories. (c) Projection of five simulated bacterial trajectories onto the $xy$-plane for $\kappa^{-1}=4$~\si{nm}. (d) The corresponding heights $h$ as a function of time for these five trajectories. The red dot is the initial position.}
\label{fig:fig7}
\end{figure}

We define the near-surface region as the area where the closest distance $h$ between the cell body and the surface is less than a critical height $h_{c}=7.0$~\si{\mu m}~\cite{Junot2022}, as illustrated in Fig.~\ref{fig:fig6}. The surface residence time $\tau$ is defined as the duration between the first and last instances that the bacteria cross the critical height $h_{c}$~\cite{Junot2022}. In subsequent simulations (Fig.~\ref{fig:fig7}), the bacteria start with an initial inclination angle of $\Psi=145^{\circ}$ and an initial height of $h=h_{c}$.

Different Debye lengths result in different stable heights $h^{*}$ of bacteria on the surface~\cite{Liu2025D}. To obtain different stable heights, we perform numerical simulations using two different Debye lengths: $\kappa^{-1}=2$~\si{nm} and $\kappa^{-1}=4$~\si{nm}. Fig.~\ref{fig:fig7} shows the projections of five different simulated trajectories onto the surface, along with the evolution of their corresponding heights over time. Figs.~\ref{fig:fig7}(a) and (b) present the simulation results for $\kappa^{-1}=2$~\si{nm}, while Figs.~\ref{fig:fig7}(c) and (d) correspond to $\kappa^{-1}=4$~\si{nm}. It can be observed from Fig.~\ref{fig:fig7} that bacteria with left-handed helical flagella tend to swim in clockwise circles. Furthermore, the surface residence time exhibits a corresponding change with the decrease in stable height $h^{*}$.

\begin{figure}
\centering
\includegraphics[width=0.50\textwidth]{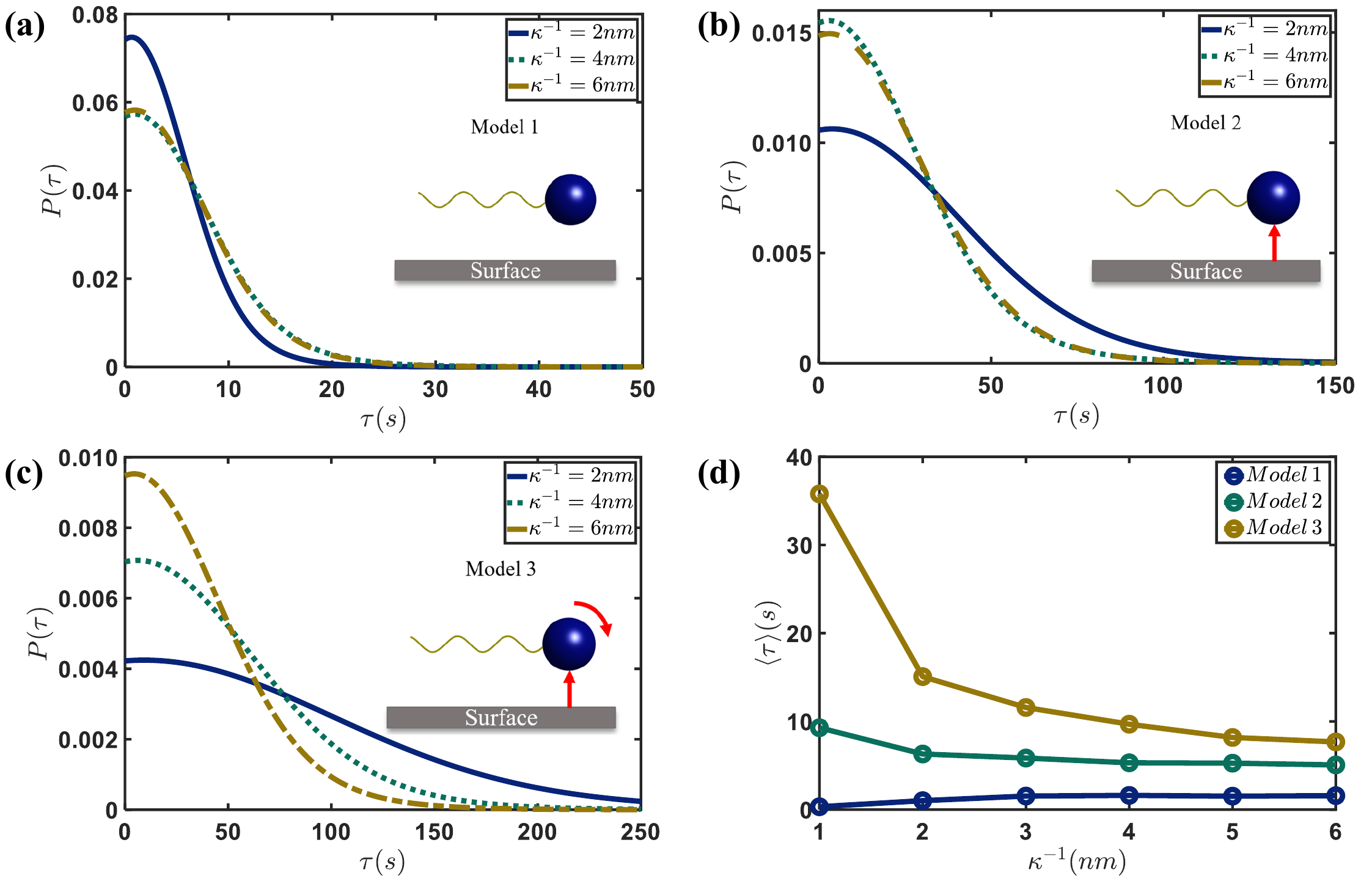}
\caption{(a)-(c) Distributions of the surface residence time for $Model\,1$, $Model\,2$, and $Model\,3$, respectively, at Debye lengths of $\kappa^{-1}=2$~\si{nm}, $\kappa^{-1}=4$~\si{nm}, and $\kappa^{-1}=6$~\si{nm}. Three different models of bacteria near a surface: $Model\,1$: without near-field HIs; $Model\,2$: neglecting the submatrix $B_{b}$ of the resistance matrix $\mathcal{R}_{b}$; $Model\,3$: including complete near-field HIs. (d) Mean surface residence time $\langle\tau\rangle$ as a function of $\kappa^{-1}$ for these three models.}
\label{fig:fig8}
\end{figure}

Fig.~\ref{fig:fig4} shows a strong correlation between stable height and Debye length, and Figs.~\ref{fig:fig7}(b) and (d) reveal that surface residence time is closely related to Debye length. However, the quantitative statistical relationship between surface residence time and Debye length remains unclear. Previous studies indicate that different Debye lengths lead to varying stable heights, ranging from a few nanometers to over $100$ nanometers~\cite{Petroff2018,Liu2025D}, which is significantly smaller than the critical height $h_{c}=7.0$~\si{\mu m}. This observation suggests that the influence of stable height on surface residence time can ultimately be attributed to near-field HIs. To quantify the influence of near-field HIs on surface residence time, we perform simulations using the parameters provided in Table~\ref{tab:table01} and three different models depicted in Fig.~\ref{fig:fig8}. $Model\,1$, which excludes near-field HIs; $Model\,2$, which omits the non-zero submatrix $B_{b}$; and $Model\,3$, which incorporates complete near-field HIs. For each model, $10^{4}$ individual bacteria are simulated for each Debye length, and the surface residence time distributions are shown in Figs.~\ref{fig:fig8}(a)-(c). The initial height of each bacterium is $h=7.0$~\si{\mu m}, and the initial inclination angle is $\Psi=145^{\circ}$.

In $Model\,1$ (Fig.~\ref{fig:fig8}(a)), an increase in Debye length leads to a slight broadening of the surface residence time distribution~\cite{Liu2025D}. $Model\,2$ (Fig.~\ref{fig:fig8}(b)), which incorporates submatrices $A_{b}$ and $C_{b}$, exhibits enhanced hydrodynamic resistance at smaller separation distances. This reduction in distance leads to a decrease in the vertical translational velocity of the cell body (Fig.~\ref{fig:fig3}). As a result, the surface residence time distribution at $\kappa^{-1}=2$~\si{nm} is observed to be broader than that for $\kappa^{-1}=4$~\si{nm} and $\kappa^{-1}=6$~\si{nm}. Incorporating complete near-field HIs, $Model\,3$ (Fig.~\ref{fig:fig8}(c)) shows that decreasing the Debye length results in broader surface residence time distributions. The main distinction between $Model\,3$ (Fig.~\ref{fig:fig8}(c)) and $Model\,2$ (Fig.~\ref{fig:fig8}(b)) is the presence of matrix term $B_{b}$. This term represents the coupling between translational velocity and torque, which drives bacteria to orient toward the surface, thereby significantly prolonging their surface residence times.

For each model and Debye length, numerical simulations are performed using $10^{4}$ individual bacteria to qualitatively determine the average residence time. The mean residence times for each model are illustrated in Fig.~\ref{fig:fig8}(d), allowing an intuitive comparison of their values across the three models. It can be observed that the submatrices $A_{b}$ and $C_{b}$ of the resistance matrix cause an increase in the mean residence time. Similarly, the submatrix $B_{b}$ significantly increases the residence time, an effect that becomes more pronounced at smaller stable heights due to the amplification of these specific interactions. When calculating the mean residence time using $Model\,1$, a reduced Debye length results in a slight decrease in the mean residence time of bacteria on the surface.

%%%%%%%%%%%%%%%%%%%%%%%%%%%%%%%%%%%%%%%%%%%%%%%%%%%%%%%%%%%%%%%%%%%%%%%%%%%%%%%%%%%%%%%%%%%%%%%%%%%%%%
\section*{Summary And Conclusions}
\label{Sect4}
%%%%%%%%%%%%%%%%%%%%%%%%%%%%%%%%%%%%%%%%%%%%%%%%%%%%%%%%%%%%%%%%%%%%%%%%%%%%%%%%%%%%%%%%%%%%%%%%%%%%%%
Near-field HIs between bacteria and a no-slip solid surface are the primary physical mechanism underlying surface entrapment, which drives their accumulation near the surface. We employ a chiral two-body model to simulate the dynamics of bacteria near the surface. The surface entrapment process can be divided into three sequential stages: approach, reorientation, and surface swimming. Bacterial motility near the surface is governed by a resistance matrix. This matrix mainly comprises three types of couplings: $A_{b}$ or force-translational velocity coupling, $C_{b}$ for torque-rotational velocity coupling and $B_{b}^{T}$ and $B_{b}$ for cross couplings. As bacteria approach the surface, the influence of $A_{b}$ and $C_{b}$ leads to increased hydrodynamic resistance, causing a decrease in vertical translational velocity $U_{\perp}$. Upon contact with the surface, $U_{\perp}$ becomes zero, effectively confining the bacteria to the surface.

However, the presence of DLVO forces prevents bacteria from contacting the surface. Under the combination of near-field HIs and the DLVO force, bacteria reach a stable fixed point $\{h^{*},\Psi^{*}\}$ in the phase plane, exhibiting circular trajectories near the surface. In particular, bacteria with left-handed helical flagella exhibit clockwise circular motion when viewed from above. During this stage, as the stable height $h^{*}$ increases, the horizontal translational velocity $U_{\parallel}$ increases correspondingly, while the vertical rotational velocity $W_{\perp}$ decreases, which together leads to a larger radius of curvature. Additionally, the radius of curvature of the bacterial circular motion is linearly correlated with the stable height $h^{*}$ for $h^{*}\ge40$~\si{nm}. Submatrices $A_{b}$ and $C_{b}$ increase the residence time of bacteria on the surface, while the submatrix $B_{b}$ shows a more pronounced effect in prolonging it.

Our research reveals that the near-field HIs between the spherical particles and a solid surface cause the translational diffusion coefficients of the particles to decrease and ultimately reach zero as the particles approach the surface. This significantly prolongs their residence time near the surface. Furthermore, previous work~\cite{Liu2025A} demonstrates that similar near-field HIs also exist between spherical particles. As two spherical particles approach each other, their translational diffusion coefficients decrease to zero, significantly increasing their timescale for separation. Building on these findings and the theory of interparticle near-field HIs, we hypothesize that these interactions promote particle crystallization when the following two conditions are met: 1. The particle size is significantly larger than the effective interaction range of DLVO forces (typically on the micrometer scale), making near-field HIs prominent. 2. The resistance matrix for the system is valid when a no-slip boundary condition is satisfied between the spherical particles. This is the focus of our next research.

%%%%%%%%%%%%%%%%%%%%%%%%%%%%%%%%%%%%%%%%%%%%%%%%%%%%%%%%%%%%%%%%%%%%%%%%%%%%%%%%%%%%%%%%%%%%%%%%%%%%%%
\section*{Author contributions}
Baopi Liu: Data curation, Formal analysis, Methodology, Software, Validation, Visualization, Writing-original draft. Lu Chen: Investigation, Methodology, Validation, Visualization. Haiqin Wang: Investigation, Methodology, Validation, Visualization. All authors reviewed and edited the manuscript.

%%%%%%%%%%%%%%%%%%%%%%%%%%%%%%%%%%%%%%%%%%%%%%%%%%%%%%%%%%%%%%%%%%%%%%%%%%%%%%%%%%%%%%%%%%%%%%%%%%%%%%
\section*{Conflicts of interest}
There are no conflicts of interest to declare.

%%%%%%%%%%%%%%%%%%%%%%%%%%%%%%%%%%%%%%%%%%%%%%%%%%%%%%%%%%%%%%%%%%%%%%%%%%%%%%%%%%%%%%%%%%%%%%%%%%%%%%
\section*{Acknowledgements}
The authors acknowledge Beijing Computational Science Research Center for providing computational support for this work.

%%%%%%%%%%%%%%%%%%%%%%%%%%%%%%%%%%%%%%%%%%%%%%%%%%%%%%%%%%%%%%%%%%%%%%%%%%%%%%%%%%%%%%%%%%%%%%%%%%%%%%
%\appendix
\section*{Appendix A: Resistance Matrix of the Flagellum}
\label{Appendix A}
%%%%%%%%%%%%%%%%%%%%%%%%%%%%%%%%%%%%%%%%%%%%%%%%%%%%%%%%%%%%%%%%%%%%%%%%%%%%%%%%%%%%%%%%%%%%%%%%%%%%%%
The rigid helical flagellum is represented by a chiral body model, and the elements of its $6\times6$ resistance matrix are~\cite{Di2011,Dvoriashyna2021,Liu2025C}
\begin{equation}
\begin{split}
X_{\parallel}^{A}=\Lambda\left[k_{\parallel}\cos^{2}\theta+k_{\perp}\sin^{2}\theta\right],
\label{eq:refnameA01}
\end{split}
\end{equation}
\begin{equation}
\begin{split}
X_{\perp}^{A}=\Lambda\left[k_{\parallel}\frac{\sin^{2}\theta}{2}+k_{\perp}\frac{1+\cos^{2}\theta}{2}\right],
\label{eq:refnameA02}
\end{split}
\end{equation}
\begin{equation}
\begin{split}
X_{\parallel}^{B}=RL\sin\theta\left(k_{\perp}-k_{\parallel}\right),
\label{eq:refnameA03}
\end{split}
\end{equation}
\begin{equation}
\begin{split}
X_{\perp}^{B}=-\frac{1}{2}RL\sin\theta\left(k_{\perp}-k_{\parallel}\right),
\label{eq:refnameA04}
\end{split}
\end{equation}
\begin{equation}
\begin{split}
X_{\parallel}^{C}=\Lambda R^{2}\left[k_{\parallel}\sin^{2}\theta+k_{\perp}\cos^{2}\theta\right],
\label{eq:refnameA05}
\end{split}
\end{equation}
\begin{equation}
\begin{split}
X_{\perp}^{C}=\Lambda\left[k_{\perp}(\frac{R^{2}}{2}+\frac{L^{2}}{12})+\left(k_{\parallel}-k_{\perp}\right)\sin^{2}\theta(\frac{R^{2}}{2\gamma^{2}}+\frac{L^{2}}{24})\right].
\label{eq:refnameA06}
\end{split}
\end{equation}
where $\gamma=2\pi R/\lambda$. The Gray and Hancock's drag coefficients are~\cite{Gray1955,Chwang1975}:
\begin{equation}
\begin{split}
&k_{\parallel}=\frac{2\pi\mu}{\ln(2\lambda/a)-1/2},\\
&k_{\perp}=\frac{4\pi\mu}{\ln(2\lambda/a)+1/2}.
\label{eq:refnameA07}
\end{split}
\end{equation}
where $\mu$ is the dynamic viscosity of the fluid and $a$ is the filament radius of the flagellum.

%%%%%%%%%%%%%%%%%%%%%%%%%%%%%%%%%%%%%%%%%%%%%%%%%%%%%%%%%%%%%%%%%%%%%%%%%%%%%%%%%%%%%%%%%%%%%%%%%%%%%%
\section*{Appendix B: Resistance Matrix of a Sphere Near a Solid Surface}
\label{Appendix B}
%%%%%%%%%%%%%%%%%%%%%%%%%%%%%%%%%%%%%%%%%%%%%%%%%%%%%%%%%%%%%%%%%%%%%%%%%%%%%%%%%%%%%%%%%%%%%%%%%%%%%%
When bacteria swim near a no-slip solid surface, the scalar functions of the resistance matrix of the spherical cell body are~\cite{Dunstan2012}:
\begin{equation}
\begin{split}
\bar{Y}_{\parallel}^{A}=\frac{Y_{\parallel}^{A}}{6\pi\mu R_{b}}=&\left(1.9963\xi-0.5332\right)\log\left(\frac{\xi}{\xi+1}\right)+2.9963\\
&-\frac{0.9689}{\xi+1}-\frac{0.5993}{(\xi+1)^{2}}-\frac{0.4691}{(\xi+1)^{3}},
\label{eq:refnameB01}
\end{split}
\end{equation}
\begin{equation}
\begin{split}
&\bar{Y}_{\perp}^{A}=\frac{Y_{\perp}^{A}}{6\pi\mu R_{b}}=\frac{2+9\xi+6\xi^{2}}{2\xi+6\xi^{2}},
\label{eq:refnameB02}
\end{split}
\end{equation}
\begin{equation}
\begin{split}
\bar{Y}^{B}=\frac{Y^{B}}{6\pi\mu R_{b}^{2}}=&\left(0.4991\xi+0.1334\right)\log\left(\frac{\xi}{\xi+1}\right)+0.4991\\
&-\frac{0.1162}{\xi+1}-\frac{0.0165}{(\xi+1)^{2}}+\frac{0.0028}{(\xi+1)^{3}},
\label{eq:refnameB03}
\end{split}
\end{equation}
\begin{equation}
\begin{split}
\bar{Y}_{\parallel}^{C}=\frac{Y_{\parallel}^{C}}{8\pi\mu R_{b}^{3}}=&-\left(0.4+0.7898\xi\right)\log\left(\frac{\xi}{\xi+1}\right)+0.2101\\
&-\frac{0.0050}{\xi+1}-\frac{0.0683}{(\xi+1)^{2}}+\frac{0.2449}{(\xi+1)^{3}},
\label{eq:refnameB04}
\end{split}
\end{equation}
\begin{equation}
\begin{split}
\bar{Y}_{\perp}^{C}=\frac{Y_{\perp}^{C}}{8\pi\mu R_{b}^{3}}=&\left[0.414\xi+0.318\frac{\xi^{2}}{\xi+1}\right]\log\left(\frac{\xi}{\xi+1}\right)+1.732\\
&-\frac{0.684}{\xi+1}+\frac{0.037}{(\xi+1)^{2}}+\frac{0.117}{(\xi+1)^{3}}.
\label{eq:refnameB05}
\end{split}
\end{equation}
where $\xi=h/R_{b}$, and $h$ is the closed distance between the cell body and the surface.

%%%%%%%%%%%%%%%%%%%%%%%%%%%%%%%%%%%%%%%%%%%%%%%%%%%%%%%%%%%%%%%%%%%%%%%%%%%%%%%%%%%%%%%%%%%%%%%%%%%%%%
\section*{Appendix C: Mobility Matrix of a Sphere Near a Solid Surface}
\label{Appendix C}
%%%%%%%%%%%%%%%%%%%%%%%%%%%%%%%%%%%%%%%%%%%%%%%%%%%%%%%%%%%%%%%%%%%%%%%%%%%%%%%%%%%%%%%%%%%%%%%%%%%%%%
\begin{figure*}
\centering
\includegraphics[width=1.00\textwidth]{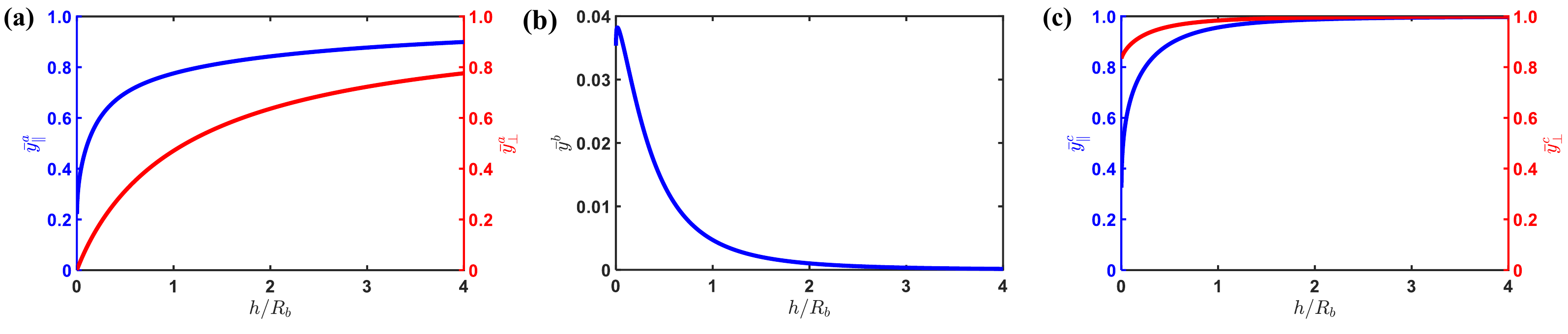}
\caption{The scalar functions (a) $\bar{y}_{\parallel}^{a}$, $\bar{y}_{\perp}^{a}$, (b) $\bar{y}^{b}$, (c) $\bar{y}_{\parallel}^{c}$ and $\bar{y}_{\perp}^{c}$ of the mobility matrix as a function of reduced height $h/R_{b}$.}
\label{fig:figC}
\end{figure*}

The scalar functions of the mobility matrix are:
\begin{equation}
\begin{split}
&\bar{y}_{\parallel}^{a}=6\pi\mu R_{b}y_{\parallel}^{a},\\
&\bar{y}_{\perp}^{a}=6\pi\mu R_{b}y_{\perp}^{a},\\
&\bar{y}^{b}=6\pi\mu R_{b}^{2}y^{b},\\
&\bar{y}_{\parallel}^{c}=8\pi\mu R_{b}^{3}y_{\parallel}^{c},\\
&\bar{y}_{\perp}^{c}=8\pi\mu R_{b}^{3}y_{\perp}^{c}.
\label{eq:refnameC01}
\end{split}
\end{equation}
The mobility matrix for a spherical cell body near a solid surface can be written as $\mathcal{M}_{b}=\mathcal{R}_{b}^{-1}$:
\begin{equation}
\begin{split}
\mathcal{M}_{b}=\left(\begin{matrix} 
y_{\parallel}^{a} & 0 & 0 & 0 & y^{b} & 0 \\
0 & y_{\parallel}^{a} & 0 & -y^{b} & 0 & 0 \\
0 & 0 & y_{\perp}^{a} & 0 & 0 & 0 \\
0 & -y^{b} & 0 & y_{\parallel}^{c} & 0 & 0 \\
y^{b} & 0 & 0 & 0 & y_{\parallel}^{c} & 0\\
0 & 0 & 0 & 0 & 0 & y_{\perp}^{c} 
\end{matrix}\right).
\label{eq:refnameC02}
\end{split}
\end{equation}
The scalar functions of the mobility matrix are shown in Fig.~\ref{fig:figC}. The translational and rotational diffusion coefficients are~\cite{Alexandre2023}:
\begin{equation}
\begin{split}
D^{T}=k_{B}T\left(\begin{matrix} 
y_{\parallel}^{a} & 0 & 0\\
0 & y_{\parallel}^{a} & 0\\
0 & 0 & y_{\perp}^{a}\\
\end{matrix}\right).
\label{eq:refnameC03}
\end{split}
\end{equation}
\begin{equation}
\begin{split}
D^{R}=k_{B}T\left(\begin{matrix} 
y_{\parallel}^{c} & 0 & 0\\
0 & y_{\parallel}^{c} & 0\\
0 & 0 & y_{\perp}^{c}\\
\end{matrix}\right).
\label{eq:refnameC04}
\end{split}
\end{equation}
Both translational and diffusion coefficients decrease with decreasing reduced height $h/R_{b}$.

%%%%%%%%%%%%%%%%%%%%%%%%%%%%%%%%%%%%%%%%%%%%%%%%%%%%%%%%%%%%%%%%%%%%%%%%%%%%%%%%%%%%%%%%%%%%%%%%%%%%%%
%%%END OF MAIN TEXT%%%
\balance
%%%REFERENCES%%%
\bibliography{rsc} %You need to replace "rsc" on this line with the name of your .bib file

@article{Tuson2013,
  title={Bacteria--surface interactions},
  author={Tuson, Hannah H and Weibel, Douglas B},
  journal={Soft matter},
  volume={9},
  number={17},
  pages={4368--4380},
  year={2013},
  publisher={Royal Society of Chemistry}
}

@article{Liu2013,
  title={Adsorption of bacteria onto layered double hydroxide particles to form biogranule-like aggregates},
  author={Liu, Jianyong and Duan, Chao and Zhou, Jizhi and Li, Xiangling and Qian, Guangren and Xu, Zhi Ping},
  journal={Applied Clay Science},
  volume={75},
  pages={39--45},
  year={2013},
  publisher={Elsevier}
}

@article{Ghosh2019,
  title={Cleaning carbohydrate impurities from lignin using Pseudomonas fluorescens},
  author={Ghosh, Tanushree and Ngo, Tri-Dung and Kumar, Aloke and Ayranci, Cagri and Tang, Tian},
  journal={Green Chemistry},
  volume={21},
  number={7},
  pages={1648--1659},
  year={2019},
  publisher={Royal Society of Chemistry}
}

@article{Monds2009,
  title={The developmental model of microbial biofilms: ten years of a paradigm up for review},
  author={Monds, Russell D and O’Toole, George A},
  journal={Trends in microbiology},
  volume={17},
  number={2},
  pages={73--87},
  year={2009},
  publisher={Elsevier}
}

@article{Carniello2018,
  title={Physico-chemistry from initial bacterial adhesion to surface-programmed biofilm growth},
  author={Carniello, Vera and Peterson, Brandon W and van der Mei, Henny C and Busscher, Henk J},
  journal={Advances in colloid and interface science},
  volume={261},
  pages={1--14},
  year={2018},
  publisher={Elsevier}
}

@article{Krsmanovic2021,
  title={Hydrodynamics and surface properties influence biofilm proliferation},
  author={Krsmanovic, Milos and Biswas, Dipankar and Ali, Hessein and Kumar, Aloke and Ghosh, Ranajay and Dickerson, Andrew K},
  journal={Advances in Colloid and Interface Science},
  volume={288},
  pages={102336},
  year={2021},
  publisher={Elsevier}
}

@article{Costerton1999,
  title={Bacterial biofilms: a common cause of persistent infections},
  author={Costerton, J William and Stewart, Philip S and Greenberg, E Peter},
  journal={science},
  volume={284},
  number={5418},
  pages={1318--1322},
  year={1999},
  publisher={American Association for the Advancement of Science}
}

@article{Donlan2001,
  title={Biofilms and device-associated infections},
  author={Donlan, Rodney M},
  journal={Emerging infectious diseases},
  volume={7},
  number={2},
  pages={277},
  year={2001}
}

@article{Chang2018,
  title={Surface topography hinders bacterial surface motility},
  author={Chang, Yow-Ren and Weeks, Eric R and Ducker, William A},
  journal={ACS applied materials \& interfaces},
  volume={10},
  number={11},
  pages={9225--9234},
  year={2018},
  publisher={ACS Publications}
}

@article{Penesyan2021,
  title={Three faces of biofilms: a microbial lifestyle, a nascent multicellular organism, and an incubator for diversity},
  author={Penesyan, Anahit and Paulsen, Ian T and Kjelleberg, Staffan and Gillings, Michael R},
  journal={npj Biofilms and Microbiomes},
  volume={7},
  number={1},
  pages={80},
  year={2021},
  publisher={Nature Publishing Group UK London}
}

@article{Lauga2016,
  title={Bacterial hydrodynamics},
  author={Lauga, Eric},
  journal={Annual Review of Fluid Mechanics},
  volume={48},
  number={1},
  pages={105--130},
  year={2016},
  publisher={Annual Reviews}
}

@article{Liu2025B,
  title={Effects of flagellar morphology on swimming performance and directional control in microswimmers},
  author={Liu, Baopi and Chen, Lu and Xu, Wenjun},
  journal={Physics of Fluids},
  volume={37},
  number={4},
  pages={041912},
  year={2025},
  publisher={AIP Publishing}
}

@article{Bianchi2017,
  title={Holographic imaging reveals the mechanism of wall entrapment in swimming bacteria},
  author={Bianchi, Silvio and Saglimbeni, Filippo and Di Leonardo, Roberto},
  journal={Physical Review X},
  volume={7},
  number={1},
  pages={011010},
  year={2017},
  publisher={APS}
}

@article{Bianchi2019,
  title={3D dynamics of bacteria wall entrapment at a water--air interface},
  author={Bianchi, Silvio and Saglimbeni, Filippo and Frangipane, Giacomo and Dell'Arciprete, Dario and Di Leonardo, Roberto},
  journal={Soft matter},
  volume={15},
  number={16},
  pages={3397--3406},
  year={2019},
  publisher={Royal Society of Chemistry}
}

@article{Berke2008,
  title={Hydrodynamic attraction of swimming microorganisms by surfaces},
  author={Berke, Allison P and Turner, Linda and Berg, Howard C and Lauga, Eric},
  journal={Physical Review Letters},
  volume={101},
  number={3},
  pages={038102},
  year={2008},
  publisher={APS}
}

@article{Li2011,
  title={Accumulation of swimming bacteria near a solid surface},
  author={Li, Guanglai and Bensson, James and Nisimova, Liana and Munger, Daniel and Mahautmr, Panrapee and Tang, Jay X and Maxey, Martin R and Brun, Yves V},
  journal={Physical Review E—Statistical, Nonlinear, and Soft Matter Physics},
  volume={84},
  number={4},
  pages={041932},
  year={2011},
  publisher={APS}
}

@article{Wu2018,
  title={Entrapment of pusher and puller bacteria near a solid surface},
  author={Wu, Kuan-Ting and Hsiao, Yi-Teng and Woon, Wei-Yen},
  journal={Physical Review E},
  volume={98},
  number={5},
  pages={052407},
  year={2018},
  publisher={APS}
}

@article{Zhang2021,
  title={An effective and efficient model of the near-field hydrodynamic interactions for active suspensions of bacteria},
  author={Zhang, Bokai and Leishangthem, Premkumar and Ding, Yang and Xu, Xinliang},
  journal={Proceedings of the National Academy of Sciences},
  volume={118},
  number={28},
  pages={e2100145118},
  year={2021},
  publisher={National Academy of Sciences}
}

@article{Sharma2003,
  title={Adhesion of Paenibacillus polymyxa on chalcopyrite and pyrite: surface thermodynamics and extended DLVO theory},
  author={Sharma, PK and Rao, K Hanumantha},
  journal={Colloids and Surfaces B: Biointerfaces},
  volume={29},
  number={1},
  pages={21--38},
  year={2003},
  publisher={Elsevier},
  doi={10.1016/S0927-7765(02)00180-7}
}

@article{Hong2012,
  title={Initial adhesion of Bacillus subtilis on soil minerals as related to their surface properties},
  author={Hong, Z and Rong, X and Cai, P and Dai, K and Liang, W and Chen, W and Huang, Q},
  journal={European Journal of Soil Science},
  volume={63},
  number={4},
  pages={457--466},
  year={2012},
  publisher={Wiley Online Library},
  doi={10.1111/j.1365-2389.2012.01460.x}
}

@article{Liu2025D,
  title={Simulation of Flagellated Bacteria Near a Solid Surface: Effects of Flagellar Morphology and Ionic Strength},
  author={Liu, Baopi and Jin, Bowen and An, Ning},
  journal={arXiv preprint arXiv:2506.19271},
  year={2025}
}

@article{Frymier1995,
  title={Three-dimensional tracking of motile bacteria near a solid planar surface},
  author={Frymier, Paul D and Ford, Roseanne M and Berg, Howard C and Cummings, Peter T},
  journal={Proceedings of the National Academy of Sciences},
  volume={92},
  number={13},
  pages={6195--6199},
  year={1995},
  doi={10.1073/pnas.92.13.6195}
}

@article{Li2008,
  title={Amplified effect of Brownian motion in bacterial near-surface swimming},
  author={Li, Guanglai and Tam, Lick-Kong and Tang, Jay X},
  journal={Proceedings of the National Academy of Sciences},
  volume={105},
  number={47},
  pages={18355--18359},
  year={2008},
  publisher={National Academy of Sciences}
}

@article{Vissers2018,
  title={Bacteria as living patchy colloids: Phenotypic heterogeneity in surface adhesion},
  author={Vissers, Teun and Brown, Aidan T and Koumakis, Nick and Dawson, Angela and Hermes, Michiel and Schwarz-Linek, Jana and Schofield, Andrew B and French, Joseph M and Koutsos, Vasileios and Arlt, Jochen and others},
  journal={Science Advances},
  volume={4},
  number={4},
  pages={eaao1170},
  year={2018},
  publisher={American Association for the Advancement of Science}
}

@article{Berne2018,
  title={Bacterial adhesion at the single-cell level},
  author={Berne, Cecile and Ellison, Courtney K and Ducret, Adrien and Brun, Yves V},
  journal={Nature Reviews Microbiology},
  volume={16},
  number={10},
  pages={616--627},
  year={2018},
  publisher={Nature Publishing Group UK London}
}

@article{Utada2014,
  title={Vibrio cholerae use pili and flagella synergistically to effect motility switching and conditional surface attachment},
  author={Utada, Andrew S and Bennett, Rachel R and Fong, Jiunn CN and Gibiansky, Maxsim L and Yildiz, Fitnat H and Golestanian, Ramin and Wong, Gerard CL},
  journal={Nature communications},
  volume={5},
  number={1},
  pages={4913},
  year={2014},
  publisher={Nature Publishing Group UK London}
}

@article{Junot2022,
  title={Run-to-tumble variability controls the surface residence times of E. coli bacteria},
  author={Junot, Gaspard and Darnige, Thierry and Lindner, Anke and Martinez, Vincent A and Arlt, Jochen and Dawson, Angela and Poon, Wilson CK and Auradou, Harold and Cl{\'e}ment, Eric},
  journal={Physical Review Letters},
  volume={128},
  number={24},
  pages={248101},
  year={2022},
  publisher={APS}
}

@article{Lauga2006,
  title={Swimming in circles: motion of bacteria near solid boundaries},
  author={Lauga, Eric and DiLuzio, Willow R and Whitesides, George M and Stone, Howard A},
  journal={Biophysical journal},
  volume={90},
  number={2},
  pages={400--412},
  year={2006},
  publisher={Elsevier}
}

@article{Khalid2020,
  title={Tuning surface topographies on biomaterials to control bacterial infection},
  author={Khalid, Saud and Gao, Ang and Wang, Guomin and Chu, Paul K and Wang, Huaiyu},
  journal={Biomaterials Science},
  volume={8},
  number={24},
  pages={6840--6857},
  year={2020},
  publisher={Royal Society of Chemistry}
}

@article{Zheng2021,
  title={Implication of surface properties, bacterial motility, and hydrodynamic conditions on bacterial surface sensing and their initial adhesion},
  author={Zheng, Sherry and Bawazir, Marwa and Dhall, Atul and Kim, Hye-Eun and He, Le and Heo, Joseph and Hwang, Geelsu},
  journal={Frontiers in Bioengineering and Biotechnology},
  volume={9},
  pages={643722},
  year={2021},
  publisher={Frontiers Media SA}
}

@article{Di2011,
  title={Swimming with an image},
  author={Di Leonardo, Roberto and Dell’Arciprete, Dario and Angelani, Luca and Iebba, Valerio},
  journal={Physical review letters},
  volume={106},
  number={3},
  pages={038101},
  year={2011},
  publisher={APS}
}

@article{Tokarova2021,
  title={Patterns of bacterial motility in microfluidics-confining environments},
  author={Tok{\'a}rov{\'a}, Viola and Sudalaiyadum Perumal, Ayyappasamy and Nayak, Monalisha and Shum, Henry and Ka{\v{s}}par, Ond{\v{r}}ej and Rajendran, Kavya and Mohammadi, Mahmood and Tremblay, Charles and Gaffney, Eamonn A and Martel, Sylvain and others},
  journal={Proceedings of the National Academy of Sciences},
  volume={118},
  number={17},
  pages={e2013925118},
  year={2021},
  publisher={National Academy of Sciences}
}

@article{Liu2025A,
  title={Effective and efficient modeling of the hydrodynamics for bacterial flagella},
  author={Liu, Baopi and Chen, Lu and Zhang, Ji},
  journal={Physics of Fluids},
  volume={37},
  number={1},
  pages={011903},
  year={2025},
  publisher={AIP Publishing}
}

@article{Petroff2015,
  title={Fast-moving bacteria self-organize into active two-dimensional crystals of rotating cells},
  author={Petroff, Alexander P and Wu, Xiao-Lun and Libchaber, Albert},
  journal={Physical review letters},
  volume={114},
  number={15},
  pages={158102},
  year={2015},
  publisher={APS}
}

@article{Sipos2015,
  title={Hydrodynamic trapping of swimming bacteria by convex walls},
  author={Sipos, Orsolya and Nagy, Katalin and Di Leonardo, R and Galajda, Peter},
  journal={Physical review letters},
  volume={114},
  number={25},
  pages={258104},
  year={2015},
  publisher={APS}
}

@article{Das2019,
  title={Transition to bound states for bacteria swimming near surfaces},
  author={Das, Debasish and Lauga, Eric},
  journal={Physical Review E},
  volume={100},
  number={4},
  pages={043117},
  year={2019},
  publisher={APS}
}

@article{Petroff2018,
  title={Nucleation of rotating crystals by Thiovulum majus bacteria},
  author={Petroff, AP and Libchaber, A},
  journal={New Journal of Physics},
  volume={20},
  number={1},
  pages={015007},
  year={2018},
  publisher={IOP Publishing}
}

@article{Molaei2014,
  title={Failed escape: solid surfaces prevent tumbling of Escherichia coli},
  author={Molaei, Mehdi and Barry, Michael and Stocker, Roman and Sheng, Jian},
  journal={Physical review letters},
  volume={113},
  number={6},
  pages={068103},
  year={2014},
  publisher={APS}
}

@article{Schaar2015,
  title={Detention times of microswimmers close to surfaces: Influence of hydrodynamic interactions and noise},
  author={Schaar, Konstantin and Z{\"o}ttl, Andreas and Stark, Holger},
  journal={Physical review letters},
  volume={115},
  number={3},
  pages={038101},
  year={2015},
  publisher={APS}
}

@article{Militaru2021,
  title={Escape dynamics of active particles in multistable potentials},
  author={Militaru, Andrei and Innerbichler, Max and Frimmer, Martin and Tebbenjohanns, Felix and Novotny, Lukas and Dellago, Christoph},
  journal={Nature Communications},
  volume={12},
  number={1},
  pages={2446},
  year={2021},
  publisher={Nature Publishing Group UK London}
}

@article{Dvoriashyna2021,
  title={Hydrodynamics and direction change of tumbling bacteria},
  author={Dvoriashyna, Mariia and Lauga, Eric},
  journal={Plos one},
  volume={16},
  number={7},
  pages={e0254551},
  year={2021},
  publisher={Public Library of Science San Francisco, CA USA},
  doi={10.1371/journal.pone.0254551}
}

@article{Liu2025C,
  title={Morphological Effects on Bacterial Brownian Motion: Validation of a Chiral Two-Body Model},
  author={Liu, Baopi and Jin, Bowen and Chen, Lu and Liu, Ning},  
  eprint={2504.05053},
  journal={arXiv preprint arXiv:2504.05053},
  archivePrefix={arXiv},
  year={2025},
  doi={10.48550/arXiv.2504.05053}
}

@article{Gray1955,
  title={The propulsion of sea-urchin spermatozoa},
  author={Gray, James and Hancock, Gregory J},
  journal={Journal of Experimental Biology},
  volume={32},
  number={4},
  pages={802--814},
  year={1955},
  publisher={The Company of Biologists Ltd},
  doi={10.1242/jeb.32.4.802}
}

@article{Chwang1975,
  title={Hydromechanics of low-Reynolds-number flow. Part 2. Singularity method for Stokes flows},
  author={Chwang, Allen T and Wu, T Yao-Tsu},
  journal={Journal of Fluid mechanics},
  volume={67},
  number={4},
  pages={787--815},
  year={1975},
  publisher={Cambridge University Press},
  doi={10.1017/S0022112075000614}
}

@article{Johnson1979,
  title={Flagellar hydrodynamics. A comparison between resistive-force theory and slender-body theory},
  author={Johnson, RE and Brokaw, CJ},
  journal={Biophysical journal},
  volume={25},
  number={1},
  pages={113--127},
  year={1979},
  publisher={Elsevier},
  doi={10.1016/S0006-3495(79)85281-9}
}

@article{Dunstan2012,
  title={A two-sphere model for bacteria swimming near solid surfaces},
  author={Dunstan, Jocelyn and Mino, Gast{\'o}n and Clement, Eric and Soto, Rodrigo},
  journal={Physics of Fluids},
  volume={24},
  number={1},
  pages={011901},
  year={2012},
  publisher={AIP Publishing}
}

@article{Rodenborn2013,
  title={Propulsion of microorganisms by a helical flagellum},
  author={Rodenborn, Bruce and Chen, Chih-Hung and Swinney, Harry L and Liu, Bin and Zhang, HP},
  journal={Proceedings of the National Academy of Sciences},
  volume={110},
  number={5},
  pages={E338--E347},
  year={2013},
  publisher={National Acad Sciences}
}

@article{Esparza2021,
  title={Dynamics of a helical swimmer crossing viscosity gradients},
  author={Esparza L{\'o}pez, Christian and Gonzalez-Gutierrez, Jorge and Solorio-Ordaz, Francisco and Lauga, Eric and Zenit, Roberto},
  journal={Physical Review Fluids},
  volume={6},
  number={8},
  pages={083102},
  year={2021},
  publisher={APS}
}

@article{Das2015,
  title={Boundaries can steer active Janus spheres},
  author={Das, Sambeeta and Garg, Astha and Campbell, Andrew I and Howse, Jonathan and Sen, Ayusman and Velegol, Darrell and Golestanian, Ramin and Ebbens, Stephen J},
  journal={Nature communications},
  volume={6},
  number={1},
  pages={8999},
  year={2015},
  publisher={Nature Publishing Group UK London}
}

@article{Vigeant2002,
  title={Reversible and irreversible adhesion of motile Escherichia coli cells analyzed by total internal reflection aqueous fluorescence microscopy},
  author={Vigeant, Margot A-S and Ford, Roseanne M and Wagner, Michael and Tamm, Lukas K},
  journal={Applied and environmental microbiology},
  volume={68},
  number={6},
  pages={2794--2801},
  year={2002},
  publisher={American Society for Microbiology}
}

@article{Lauga2009,
  title={The hydrodynamics of swimming microorganisms},
  author={Lauga, Eric and Powers, Thomas R},
  journal={Reports on progress in physics},
  volume={72},
  number={9},
  pages={096601},
  year={2009},
  publisher={IOP Publishing}
}

@article{Shum2010,
  title={Modelling bacterial behaviour close to a no-slip plane boundary: the influence of bacterial geometry},
  author={Shum, Henry and Gaffney, Eamonn A and Smith, David J},
  journal={Proceedings of the Royal Society A: Mathematical, Physical and Engineering Sciences},
  volume={466},
  number={2118},
  pages={1725--1748},
  year={2010},
  publisher={The Royal Society Publishing},
  doi={10.1098/rspa.2009.0520}
}

@article{Shum2015,
  title={Hydrodynamic analysis of flagellated bacteria swimming near one and between two no-slip plane boundaries},
  author={Shum, Henry and Gaffney, Eamonn A},
  journal={Physical Review E},
  volume={91},
  number={3},
  pages={033012},
  year={2015},
  publisher={APS},
  doi={10.1103/PhysRevE.91.033012}
}

@article{Liu2024,
  title={Speed-dependent bacterial surface swimming},
  author={Liu, Qiuqian and Zhang, Chi and Zhang, Rongjing and Yuan, Junhua},
  journal={Applied and Environmental Microbiology},
  volume={90},
  number={6},
  pages={e00508--24},
  year={2024},
  publisher={American Society for Microbiology 1752 N St., NW, Washington, DC}
}

@article{Alexandre2023,
  title={Non-Gaussian diffusion near surfaces},
  author={Alexandre, Arthur and Lavaud, Maxime and Fares, Nicolas and Millan, Elodie and Louyer, Yann and Salez, Thomas and Amarouchene, Yacine and Gu{\'e}rin, Thomas and Dean, David S},
  journal={Physical review letters},
  volume={130},
  number={7},
  pages={077101},
  year={2023},
  publisher={APS}
}
\bibliographystyle{rsc} %the RSC's .bst file

\end{document}